\documentclass[fleqn,usenatbib]{mnras}

\usepackage{newtxtext,newtxmath}

\usepackage{bm}

\usepackage[T1]{fontenc}

\usepackage{pifont}
\usepackage{tikz}
\usepackage{ulem}

\DeclareRobustCommand{\VAN}[3]{#2}
\let\VANthebibliography\thebibliography
\def\thebibliography{\DeclareRobustCommand{\VAN}[3]{##3}\VANthebibliography}

\usepackage{graphicx}	
\usepackage{amsmath}	
\usepackage{newtxtext,newtxmath}

\title[Planetary nurseries]{Planetary nurseries: vortices formed at smooth viscosity transition}

\author[Zs. Regály, K. Kadam \& D. Tarczay-Nehéz]{
Zs. Regály$^{1,2}$\thanks{E-mail: regaly@konkoly.hu}, K. Kadam$^{3}$ and D. Tarczay-Nehéz$^{1,2,4}$\\
$^{1}$Konkoly Observatory, Research Centre for Astronomy and Earth Science, Konkoly-Thege Mikl\'os 15-17, 1121, Budapest, Hungary\\
$^{2}$CSFK, MTA Centre of Excellence, Budapest, Konkoly Thege Miklós út 15-17., H-1121, Hungary\\
$^3$University of Western Ontario, Department of Physics and Astronomy, London, ON, N6A 3K7, Canada\\
$^{4}$ MTA CSFK Lendület Near-Field Cosmology Research Group  \\
}

\date{Accepted XXX. Received YYY; in original form ZZZ}

\pubyear{2020}

\begin{document}
\label{firstpage}
\pagerange{\pageref{firstpage}--\pageref{lastpage}}
\maketitle

\begin{abstract}
Excitation of Rossby wave instability and development of a large-scale vortex at the outer dead zone edge of protoplanetary discs is one of the leading theories that explains horseshoe-like brightness distribution in transition discs. Formation of such vortices requires a relatively sharp viscosity transition. Detailed modelling, however, indicates that viscosity transitions at the outer edge of the dead zone is relatively smooth. In this study, we present 2D global, non-isothermal, gas-dust coupled hydrodynamic simulations to investigate the possibility of vortex excitation at smooth viscosity transitions. Our models are based on a recently postulated scenario, wherein the recombination of charged particles on the surface of dust grains results in reduced ionisation fraction and in turn the turbulence due to magnetorotational instability. Thus, the $\alpha$-parameter for the disc viscosity depends on the local dust-to-gas mass ratio. We found that the smooth viscosity transitions at the outer edge of the dead zone can become Rossby unstable and form vortices. A single large-scale vortex develops if the dust content of the disc is well coupled to the gas, however, multiple small-scale vortices ensue for the case of less coupled dust. As both type of vortices are trapped at the dead zone outer edge, they provide suﬃcient time for dust growth. The solid content collected by the vortices can exceed several hundred Earth masses, while the dust-to-gas density ratio within often exceeds unity. Thus, such vortices function as planetary nurseries within the disc, providing ideal sites for formation of planetesimals and eventually planetary systems.
\end{abstract}




\section{Introduction}

Brightness asymmetries of transitional discs (an evolved phase of protoplanetary discs, \citealp{Strometal1989,Skrutskieetal1990}) observed in the past decades are a tell-tale sign of large-scale vortices, (see, e.g., \citealp{Brownetal2009,Andrewsetal2009,Hughesetal2009, Isellaetal2010,Andrewsetal2011,Mathewsetal2012,Tangetal2012,Fukagawaetal2013,Casassusetal2013,vanderMareletal2013,Perezetal2014,Hashimotoetal2015,Casassusetal2015,Wrightetal2015,Momoseetal2015,Marinoetal2015,Perezetal2018, Casassusetal2019,FrancisvanderMarel2020,Facchinietal2020,Boehleretal2021,Kurtovicetal2021,Vargaetal2021}).
As of today, we know several competing theories that can explain the origin of a large-scale vortex in protoplanetary discs:
the baroclinic instability \citep{KlahrBodenheimer2003,LyraKlahr2011,Raettigetal2013}; the convective overstability 
\citep{KlahrHubbard2014,Lyra2014};
the vertical shear instability \citep{Richardetal2016} or the zombie vortex instability \citep{Marcusetal2015}; and finally Rossby wave instability (see, a review in \citealp{LovelaceRomanova2014} and references therein).

In this study, we focus on the phenomenon of Rossby wave instability (RWI) in protoplanetary discs. RWI can be excited at a steep pressure gradient, which can form at the edges of a gap opened by an embedded planet \citep{Lietal2005} or at the regions where the disc viscosity changes considerably \citep{Lovelaceetal1999,Lietal2000}. 
\citet{Gammie1996} proposed the existence of a radially extended region in the disc midplane called the ``dead zone", where the accretion is widely reduced. 
The temperature at a distance of a few au from the central star is not high enough to collisionally ionise the gas, while the gas is dense enough to shield the external sources of ionisation such as the Galactic cosmic rays.
As a consequence of the low ionisation rate, the disc midplane is magnetically dead, resulting in low turbulence generation via magnetorotational instability (MRI).
Thus, the viscous accretion takes place only in the tenuous disc atmosphere. 
At the inner or outer edge of such a dead zone, the mismatch in the accretion rate due to the viscosity transition can result in steep pressure bumps, which become Rossby unstable and develop anticyclonic vortices.

In the past ten years, several studies have been dedicated to explaining the arc-like asymmetries observed in transitional discs (see \citealp{Andrews2020} and references therein).
According to the thorough analysis of \citet{Regalyetal2017} both the gap edge and dead zone edge scenarios can explain the observations. 
A relatively concentrated ($\lesssim 90^\circ$) strong azimuthal brightness asymmetry favours the gap edge, while an azimuthally extended brightness asymmetry ($\gtrsim 180^\circ$) dead zone edge formation scenario, respectively.
Note that both formation scenarios require a relatively low disc viscosity to maintain the vortex for enough long time to be observable.
The disc dead zone provides a low viscosity, however, the disc viscosity is presumably high at the gap edge due to the low gas density in the outer disc. 
Moreover, as recently shown by \citet{Hammeretal2017} the large-scale vortex formation at a planetary gap edge requires a fast planet growth.
A caveat for the dead zone edge scenario is that if the width of the viscosity transition at its outer edge is wider than two times the local pressure scale height, no RWI excitation occurs \citep{Lyraetal2009,Regalyetal2012}.
However, detailed modelling of disc suggests that the width of the dead zone outer edge is wider than the above criteria \citep{Dzyurkevichetal2013, Kadam19}.
Note, however, that a smooth change in gas resistivity does not imply a smooth transition in turbulent stress, for which case a large-scale vortex can form although the resistivity transition is smooth \citep{Lyraetal2015}.
Thus, further investigations of the dead zone edge scenario are warranted.

The Shakura–Sunyaev $\alpha$ viscosity model postulates a simple scaling relation between the gas viscosity and the efficiency (described by the magnitude of $\alpha$ parameter) of angular momentum transport \citep{ShakuraSunyaev1973}.
In summary, an accretion disc coupled dynamically to a weak magnetic field is subject to MRI turbulence and the
resulting in outward angular momentum transport can be described as a diffusive process with an effective $\alpha$-viscosity \citep{BalbusHawley1991}. 
The sustenance of MRI requires a sufficient level of ionisation within the disc, however, small-sized dust grains tend to adsorb electrons and ions.
This can strongly reduce the conductivity of the gas, which can inhibit MRI in the vicinity of dust particles \citep{Sanoetal2000,IlgnerNelson2006,Okuzumi2009,Johansenetal2011}

Based on the implicit assumption that the disc viscosity depends on the ionisation fraction controlled by the amount of dust concentration, 
\citet{DullemondPenzlin2018} constructed a simple parametric-$\alpha$ viscosity model.  
With 1D perturbation analysis, they demonstrated that such a parametric-$\alpha$ model leads to ring formation via viscous ring-instability (VRI).
Note that \citet{Wunschetal2005} have also shown that the disc dead zone can be split into rings due to the positive feedback between the thickness of the dead zone and the mass accumulation rate.
The positive feedback loop that leads to VRI is as follows:
1) an initial increase in the dust density reduces the local MRI viscosity, which in turn leads to the accumulation of gas in its vicinity due to the mismatch in the mass transfer rate; 
2) as the dust particles drift towards the resulting pressure maxima, they amplify the initial perturbation, leading to spontaneous formation of concentric rings.
Thus, VRI is suggested to be a potential mechanism behind the grand design ring structures often observed in the sub-millimetre emissions of protoplanetary discs \citep{Andrewsetal2018,Dullemondetal2018}.
However, 2D hydrodynamic simulations of \citet{Regalyetal2021} revealed that such rings formed via VRI are typically Rossby unstable and tend to fragment into a cascade of small-scale vortices.
They also show that grand design rings can only be stable patterns (Rossby stable) if the minimum of the reduced disc viscosity is 90 per cent that of global viscosity.

Considering the above-mentioned issues, we study RWI excitation for a smooth dead zone edge, where the width of the viscosity transition is above two times the local pressure scale height. 
We combined the standard dead zone edge $\alpha$ model (see, e.g., \citealp{Regalyetal2012}), with a parametric–$\alpha$ viscosity prescription wherein the disc viscosity depends on the local enhancement or depression of the dust content.
We present 2D non-isothermal, gas-dust coupled hydrodynamic simulations of protoplanetary discs that propose a new possibility of RWI excitation at smooth viscosity transitions.

The paper is structured as follows. 
In Section~2, the applied numerical methods for modelling RWI excitation in a dusty protoplanetary disc assuming four different parametric-$\alpha$ prescriptions are presented. 
Results and discussions are presented in Section~3 and Section ~4, respectively. 
The paper closes with our conclusion and some concluding remarks in Section~5.

\section{Numerical Methods}

\label{sec:maths} 

\subsection{Coupled gas-dust Hydrodynamics}

To study the excitation of Rossby wave instability and subsequent vortex evolution, we run coupled gas-dust hydrodynamic global disc simulations with a adiabatic thermodynamics. The dust is assumed to be a pressureless fluid. We use {\small GFARGO2} for this investigation, which is our extension to {\small GFARGO}, a GPU supported version of the {\small FARGO} code \citep{Masset2000}. The dynamics of gas and solid components are described by the following equations:
\begin{flalign}
&  \frac{\partial \Sigma_\mathrm{g}}{\partial t}+\nabla \cdot (\Sigma_\mathrm{g} {\bm{v}})=0, \label{eq:cont}\\
&  \frac{\partial \bm{v}}{\partial t}+(\bm{v} \cdot \nabla)\bm{v}=-\frac{1}{\Sigma_\mathrm{g}} \nabla P-\nabla \Phi+\nabla\cdot \bm{T}-\frac{1}{\Sigma_\mathrm{g}}\bm{f}_\mathrm{drag}, \label{eq:NS}\\
&  \frac{\partial{e}}{\partial t}+\nabla \cdot (e \bm{v})=-P\nabla{\cdot \bm{v}}+Q_\nu+Q_\pm,\label{eq:ENERG}\\
&  \frac{\partial \Sigma_\mathrm{d}}{\partial t}+\nabla \cdot \Sigma_\mathrm{d} \bm{u}=-\nabla\cdot\bm{j}, \label{eq:contd}\\
&   \frac{\partial \bm{u}}{\partial t}+(\bm{u} \cdot \nabla)\bm{u}=- \nabla\Phi+\frac{1}{\Sigma_\mathrm{d}}\bm{f}_\mathrm{drag}-\bm{u}\nabla\cdot\bm{j}, \label{eq:NSd}
\end{flalign} 
where $\Sigma_\mathrm{g}$, $\Sigma_\mathrm{d}$, and $\bm{v}$, $\bm{u}$ are the surface mass densities and velocities of gas and solid (being either dust particles or pebbles), respectively. $e$ is the thermal energy density of the gas (per surface area). 
$\bm{T}$ is the viscous stress tensor of the gas, see details in Section~\ref{sec:alpha}. ${\bm f}_\mathrm{drag}$ and ${\bm j}$ are the drag force exerted by the gas on the dust and the diffusive flux, respectively, see details in Section~\ref{sec:dust}. 
Note that the last term in Equation~(\ref{eq:NSd}) accounts for the microscopical stochastic kicks from the turbulent gas onto the solid component \citep{Bentez-Llambayetal2019,Weberetal2019}.
Emphasise that the dust back-reaction is taken into account as Equation~(\ref{eq:NS}) includes a term proportional to the drag force. The adiabatic gas pressure is given as 
\begin{equation}
P=(\gamma-1)e,
\end{equation}
where $\gamma=1.4$ is the adiabatic index of gas assumed to be H$^2$ molecule. 
The treatment of thermodynamics can have significant impact on vortex evolution in a \cite{ShakuraSunyaev1973} $\alpha$-disc.
As the local sound speed is connected to the internal thermal energy ($c_\mathrm{s} = \sqrt{\gamma P /\Sigma}$), the viscosity of the gas is different from that is in the locally isothermal case. 
This affects the onset of vortex formation, so that they tend to occur earlier, while vortex strength and lifetime are also affected \citep{TarczayNehezregaly2020}.

The total gravitational potential of the disc in Equation~(\ref{eq:NS}) is
\begin{equation}
\Phi(r,\phi)=-G\frac{M_*}{r}+\Phi_\mathrm{ind}(r,\phi),
\label{eq:phi_tot}
\end{equation} 
where the first term is the gravitational potential of the star in a given cell with radial distance $r$. Since the equations are solved in the cylindrical coordinate system centred on the star, Equation~(\ref{eq:phi_tot}) includes the so-called indirect potential, $\Phi_\mathrm{ind}(r,\phi)$ arising due to the displacement of the barycentre of the system caused by any disc non-axisymmetry (see its importance in, e.g., \citealp{MittalChiang2015,ZhuBaruteau2016, RegalyVorobyov2017a}). The indirect potential is calculated as
\begin{equation}
\Phi_\mathrm{ind}(r,\phi) =  r\cdot G\int { 
\mathrm{d}m(\bm{r}^\prime) \over {\bm r}^{\prime 3} } {\bm r}^\prime,
\label{starAccel}
\end{equation} 
which in the cylindrical coordinate system can be given as
\begin{eqnarray}
\Phi_\mathrm{ind}(r_j,\phi_k) & = &  r_{j} \cos(\phi_k) \sum_{j',k'}G {{\bm m}_{j',k'} \over r_j'^2 }\cos(\phi_{k'}) \nonumber \\
& & + \sin(\phi_k) \sum_{j',k'}G {{\bm m}_{j',k'} \over r_j'^2 }\sin(\phi_{k'}),
\label{eq:phi_ind}
\end{eqnarray}
where $m_{j,k}$ and  $r_j$, $\phi_k$, are the mass and cylindrical coordinates of the grid cell $j,k$. 
As one can see, the disc self-gravity is neglected in this investigation, however, its effect on RWI and subsequent vortex formation might be crucial (see, e.g., \citealp{RegalyVorobyov2017b,TarczayNehezetal2022} and references therein). 

The heating due to the viscous stresses, $Q_\nu$, are taken into to account according to \citet{DAngeloetal2003}. Heating and cooling mechanisms such as stellar and background irradiation are modelled implicitly via $Q_\pm$. We used $\beta$-cooling/heating prescription of \citet{LesandLin2015} 
\begin{equation}
\label{eq:qparam}
    Q_\pm = \frac{1}{\tau_c}\left ( e - e^{(0)}\frac{\Sigma}{\Sigma^{(0)}} \right ),
\end{equation}
\noindent where $\tau_c$ is the cooling time connected with the $\beta$-parameter as
\begin{equation}
\label{eq:tauc}
 \tau_c = \frac{\beta}{\Omega}, 
\end{equation}
where $\beta$ is a measure of the cooling time in terms of local dynamical timescale, $\Omega^{-1}$.
To let the gas release/gain its internal energy we assume $\beta=10$, i.e., the e-folding timescale of disc cooling/heating to the initial temperature is ten orbits at all distances (see its importance on vortex evolution in \citealp{TarczayNehezregaly2020, Rometschetal2021}). 
Note that the heating due to dust-gas friction is not included in our model.
In Equation (\ref{eq:qparam}), $\Sigma^{(0)}$ and $e^{(0)}$ correspond to the initial density and energy state of the disc.

\subsection{Viscosity prescription}
\label{sec:alpha}

The viscous stress tensor of the gas, $\bm{T}$ in Equation~(\ref{eq:NS}) is calculated as
\begin{equation}
    \bm{T} = \nu\left(\nabla \bm{v} + \nabla \bm{v}^T -\frac{2}{3}\nabla\cdot\bm{v}\bm{I}\right),
\end{equation}
where $\nu$ is the vertically integrated effective viscosity of the disc
whose components in polar coordinates are calculated according to \cite{Masset2002} and ${\bm I}$ is the identity matrix.
We use $\alpha$ prescription of \citet{ShakuraSunyaev1973} for the disc effective viscosity. In this case $\nu=\alpha c_\mathrm{s}^2/\Omega$, where the sound speed is 
\begin{equation}
c_\mathrm{s}=\sqrt{\frac{\gamma(\gamma-1)e}{\Sigma_\mathrm{g}}}.
\end{equation}
$\alpha$ characterising the strength of magneto-rotational instability depends on local dust-to-gas mass ratio. To model viscosity transition at disc dead zone the background viscosity, $\alpha_\mathrm{bg}$ is defined as 
\begin{equation}
\alpha_\mathrm{bg}=1-\frac{1}{2}\left(1-\alpha_\mathrm{mod}\right)\left[1-\tanh\left(\frac{R-R_\mathrm{dze}}{\Delta R_\mathrm{dze}}\right)\right],
\label{eq:deltaalpha}
\end{equation}
The outer edge of the disc dead zone is set to $R_\mathrm{dze}=1.5$. Excitation of RWI requires a sharp viscosity transition ($\Delta R_\mathrm{dze}\leq2H_\mathrm{dze}$, where $H_\mathrm{dze}=R_\mathrm{dze}h$ is the disc scale height at the viscosity reduction),  in the $\alpha$-prescription \citep{Lyraetal2009b,Regalyetal2012}, which is sharper than it is expected to form at the outer dead zone edge \citep{Dzyurkevichetal2013}. In our standard models, we assume that $\Delta R_\mathrm{dze}=0.2$, which corresponds to $\Delta R_\mathrm{dze}=(2+2/3)H_\mathrm{dze}$. To extend our investigation, we run models with $\Delta R_\mathrm{dze}=4H_\mathrm{dze}$ for certain cases. 
Note that the total width of the viscosity transition given by Equation\,(\ref{eq:deltaalpha}) is about $2\Delta R_\mathrm{dze}$, which corresponds to 0.4 and 0.6 in the standard and extended scenarios, respectively.

Additionally, the global viscosity is modified such that a change in the dust or gas concentration alters the viscosity. According to \citet{DullemondPenzlin2018} a parametric $\alpha$-prescription can be given as 
\begin{equation}
\alpha=\alpha_\mathrm{bg}\left(\frac{\Sigma_\mathrm{d}}{\Sigma^{(0)}_\mathrm{d}}\right)^{\phi_\mathrm{d}}\left(\frac{\Sigma_\mathrm{g}}{\Sigma^{(0)}_\mathrm{g}}\right)^{\phi_\mathrm{g}},
\label{eq:alpha}
\end{equation}
where $\Sigma^{(0)}_\mathrm{g}$ and $\Sigma^{(0)}_\mathrm{d}$ are the initial gas and dust densities. Values of $\alpha$ are limited between the background viscosity, $\alpha_\mathrm{bg}=10^{-2}$, and the MRI inactive viscosity, $\alpha_\mathrm{dead}=10^{-4}$. The assumption of minimum effective viscosity is consistent with the vertical shear instability generated viscosity\citep{StollKley2014}. With regards to the values for $\phi_\mathrm{d}$ and $\phi_\mathrm{g}$ we investigated four cases, see Table\,\ref{tbl:cases} 

\begin{table}
	\centering
	\caption{Investigated parametric–$\alpha$ models.}
	\label{tbl:cases}
	\begin{tabular}{lcc} 
		\hline
		model & $\phi_\mathrm{d}$ & $\phi\mathrm{g}$ \\
		\hline
		case A &  0 & -1 \\
		case B & -1 &  0 \\
		case C & -1 &  1 \\
		case D & -2 &  1 \\
		case E & -1 &  -1 \\
		\hline
	\end{tabular}
\end{table}

\subsection{Dust handling}
\label{sec:dust}

The turbulent diffusion of solid is modelled by the so-called gradient diffusion approximation  \citep{MorfillVoelk1984,Dubrulleetal1995,TakeuchiLin2002}, in which the diffusive flux, $\bm{j}$, is given as 
\begin{equation}
\bm{j}=-D\left(\Sigma_\mathrm{g}+\Sigma_\mathrm{d}\right)\nabla\frac{\Sigma_\mathrm{d}}{\Sigma_\mathrm{g}+\Sigma_\mathrm{d}},
\end{equation}
where the diffusion coefficient of solids is defined  according to \citet{YoudinLithwick2007} as 
\begin{equation}
D=\frac{\nu}{(1+\mathrm{St}^2)}.
    \label{eq:diffusion}
\end{equation}

Equations~(\ref{eq:NS}) and (\ref{eq:NSd}) are solved by a two-step method. First, the source term, i.e. the right-hand sides are calculated then it is followed by the conventional advection calculation. For the source term, we use a fully implicit scheme (see details in \citealp{Stoyanovskayaetal2018}). With this scheme, the effect of aerodynamic drag can be modeled for dust species that have stopping time that is much smaller than the time-step ($\tau_\mathrm{s}\ll\Delta t$). 
For pebbles that have large stopping time ($\tau_\mathrm{s}\gg\Delta t$), the method described is applicable as long as crossing orbits are not important for the dynamics. 

The drag force exerted by the gas on the dust is calculated as
\begin{equation}
    \bm{f}_\mathrm{drag}=\frac{\bm{v}-\bm{u}}{\tau_\mathrm{s}},
    \label{eq:fdrag}
\end{equation}
where $\tau_\mathrm{s}=\mathrm{St}/\Omega$ is the stopping time  and St is the Stokes number of the given solid species.
For simplicity, we assume that the solid has a fix Stokes number. 
We modeled five different solid species, whose Stokes numbers are in the range  $10^{-5}\leq\mathrm{St}\leq10^{-1}$.

The dust feedback can destroy a vortex when the local dust-to-gas mass ratio approaches 0.3–0.5, independent of the dust size \citep{Crnkovic-Rubsamenetal2015}. A vortex may be destroyed when this ratio is as low as 0.1 \citep{Johansenetal2004}.
Moreover, dust grains concentrate differentially inside the vortex and affect the gas dynamics in different ways, including vortex morphology \citep{Mirandaetal2016}.
Because of these findings and the possibility that the dust-to-gas density ratio can reach unity near the vortex eye, taking into account the dust feedback via the drag term in Eq.~(\ref{eq:NS}) is essential.

\subsection{Initial and boundary conditions}

The numerical domain extension is such that $0.5\leq R\leq5$ with $N_\mathrm{R}=512$ logarithmically distributed radial and $N_\phi=1024$ equidistant azimuthal cells. 
Both the inner and outer edge the velocity components and density of gas are damped to the initial value according to the method described in \citet{deValBorroetal2006}. 

The initial distribution of gas is set to a power-law function of $R$ as 
\begin{equation}
\Sigma_\mathrm{g}^{(0)}=\Sigma_{0} R^{-1}.
\end{equation}
The disc self-gravity is neglected, therefore $\Sigma_0$ is a free parameter in our model.

The initial dust-to-gas mass ratio is set uniformly  $\Sigma^{(0)}_\mathrm{d}/\Sigma^{(0)}_\mathrm{g}=10^{-2}$. Since the dust back-reaction is taken into account, the radial and azimuthal velocity components of dust at the initial state are taken form \citet{Garateetal2020}

Assuming that the disc initial temperature is proportinal to $R^{-1}$, the initial energy density is given as
\begin{equation}
e^{(0)}=\frac{1}{\gamma-1}\Sigma^{(0)}_\mathrm{g}h_g^2R^{-1},
\end{equation}
where $h_\mathrm{g}$ is the aspect ratio of the gaseous disc, which results in $H_g=h_gR$ local pressure scale height.

\subsection{Methods of analysis}
\label{subsec:moa}

\begin{figure*}
    \centering
    \includegraphics[width=\textwidth]{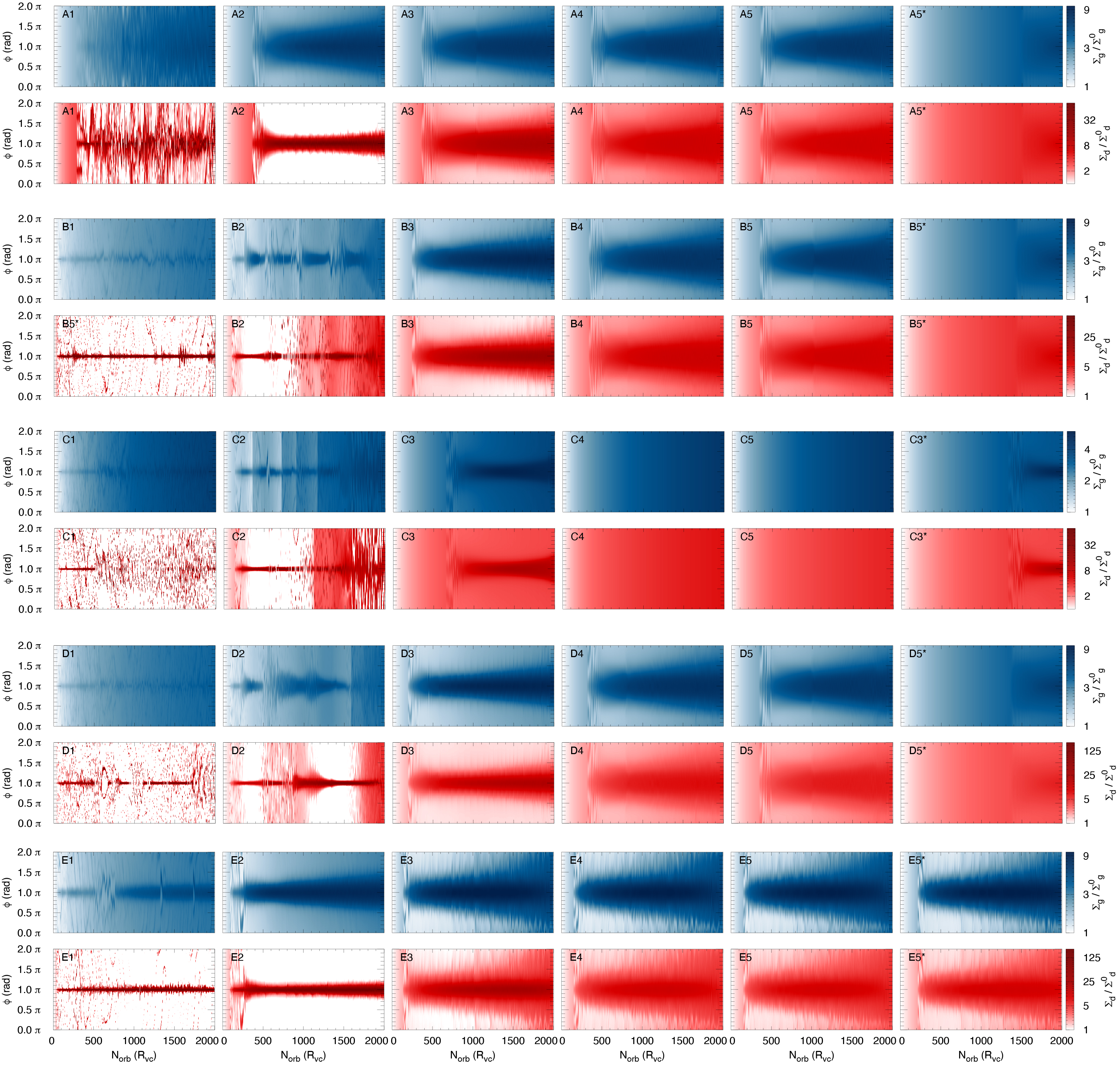}
    \caption{The time evolution of the $\delta \Sigma$ profiles in all models listed on Table 1. Blue and red panels correspond to the gas and dust azimuthal profiles, respectively. Time is measured by the number of orbits at the dead zone edge. Three patterns cam be recognised both in gas and dust profile evolution: 1) absent of any variation in the profile, because no RWI is excited; 2) trombone-like pattern, in which case a single large-scale vortex develops whose azimuthal size increases with time; 3) scattered patterns, in which case several small scale vortices develop. In the latter case the scattering occurs because the algorithm used to generate the plots are inadequate for multiple vortices, see more details in the text.}
    \label{fig:tromb}
\end{figure*}

In this section, we describe the methods for analysis and characterisation of vortices, which are used to gain insight into their formation and evolution within the disc.
The first method is identical to the one applied in \citet{RegalyVorobyov2017a}. This method is mainly suitable for the scenario of formation of large-scale vortices and is summarised as follows. 
First, the gas surface density on a 2D polar grid is normalised by the initial distribution ($\Sigma^{(0)}_{g}$). 
Then the normalised surface density is averaged radially, taking into account rings having $\pm~3$H local pressure radial distance centred on the maximum density.
Finally, the radially averaged azimuthal profiles, $\delta\Sigma$, generated from each frame throughout the simulation are displayed such that the magnitude of the profile is colour-coded. 
As a result, the time evolution of the vortex can be inferred.
The same process was applied for the dust component as well, which is normalised with respect to $\Sigma_\mathrm{d}^{(0)}$.
Results of the above-described analysis are shown in Fig.~\ref{fig:tromb}, where the time evolution (time is measured in units of the Keplerian orbit at the distance of $R=1$, which coincides with the vortex centre) of $\delta \Sigma$-profiles of gas (blue) and dust (red) are displayed.
Although this method was applied to all models, it primarily gives valuable information in the scenario of a single large-scale vortex. Details on the large-scale vortex evolution can be seen in Section\,\ref{sec:largescale-vtx}

In order to identify multiple small-scale vortices in the disc and quantify their properties, we use another method, which is identical to that described in \cite{Regalyetal2021}. 
Since a vortex forms a local maximum in gas pressure, it rapidly accumulates dust from its surroundings. 
This creates an azimuthal asymmetry in the surface distribution of dust-to-gas mass ratio ($m_{\rm d}/m_{\rm g}$).
We thus used the local maxima in the field of $(m_{\rm d}/m_{\rm g}) / \langle m_{\rm d}/ m_{\rm g} \rangle$ for detecting the location of all vortices in the disc at a given time.
As the denominator can get unintentionally large in the presence of a vortex, it was azimuthally averaged at the given radius only over the lowest half of the values.
A region extending 10 grid cells in both radial and azimuthal directions was assumed to have a local maximum in this field if the difference between the maximum and the minimum value in this region exceeded a certain threshold.
The optimal value of this threshold depends on the properties of the vortices formed in the model and needed to be set manually.
If any two maxima occurred in close proximity, they were considered to be multiple detection of a single vortex. 
The associated vortex was then assumed to be the located at the larger $\Sigma_{\rm d}$.
The closely spaced maxima were eliminated if the distance between them was less than $0.2\times R$ au and the radial separation $\Delta R < 0.05$ au.  
In order to calculate aggregate properties of the vortices, it is necessary to find the area occupied by the vortices.
The shape or area of a vortex was considered to be an ellipse in cylindrical coordinate system, centred at its location \citep[see more in e.g.][]{Kida1981,Chavanis2000}.
The semi-major and semi-minor axes of such an ellipse were empirically assumed to be a function of the radial position, such that
$a_{\theta} = 0.24 (R/5)^{0.5}$ and $a_{R} = 0.03 (5/R)^{0.8} $, where $R$ is in the units of au.
The shape thus defined was congruent with the Rossby vortices formed in the disc and the area was conservative so that all of the dust accumulated inside a typical vortex was counted.
The vortices formed close to the inner disc radius, i.e., less than $R=0.59$, were rejected because of the boundary effects.
Note that the vortex shape as described here was only used for characterising certain properties, e.g., dust mass contained within a vortex. 
These properties are not overly sensitive to the parameters chosen for vortex detection or characterisation, and we find this method to be sufficiently accurate for our purpose.

We measured the dust and gas mass ($m_\mathrm{g}$ and $m_\mathrm{d}$) within the area of the vortex. 
Two possible disc masses were investigated for each model by re-scaling the gas surface density at 1 au ($\Sigma_{\rm g, 0}$) to $2000$ and $4500$ ${\rm g\, cm^{-2}}$, which covers the range for the canonical estimates for MMSN \citep{Adams2010}. 
For multiple vortex scenarios, the largest, as well as an average value for $m_\mathrm{g}$ and $m_\mathrm{d}$, were derived.
The vortex centre was assumed to be the location of highest normalised dust-to-gas ratio.
The dust-to-gas volumetric density ratio is calculated by assuming vertical equilibrium
\begin{equation}
    \frac{\rho_\mathrm{d}}{\rho_\mathrm{g}}={\frac{1}{\sqrt{2\pi}}\Sigma_\mathrm{d}\frac{1}{H_\mathrm{d}}}/{\frac{1}{\sqrt{2\pi}}\Sigma_\mathrm{g}\frac{1}{H_\mathrm{g}}},
\end{equation}
where the gas pressure scale height is
\begin{equation}
    H_{\mathrm{g}}=\frac{c_\mathrm{s}}{\Omega_\mathrm{K}}.
\end{equation}
To calculate the dust scale height, we assume size dependent vertical sedimentation for the dust, in which case,
\begin{equation}
    H_{\mathrm{d}}=H_\mathrm{g}\sqrt{\frac{\alpha}{\mathrm{St}+\alpha}}.
\end{equation}
The fragmentation size, $a_{\rm frag}$, is the maximum size of solid constituent that can be reached due to the dust growth process, before the grown particles are destroyed via mutual collision. 
We calculate the dust grain fragmentation according to \cite{Birnstieletal2012} as
\begin{equation}
a_{\rm frag} = \frac{2 \Sigma_{\rm g} v_{\rm frag}^2}{ 3 \pi \rho_{\rm s} \alpha c_{\rm s}^2}, 
\label{eq:afrag}
\end{equation}
with the typical assumptions of fragmentation velocity, $v_{\rm frag}=10$~m~s$^{-1}$, and the internal density of the dust aggregate, $\rho_{\rm s} = 1.6~\mathrm{g~cm}^{-3}$.
The value of $\alpha$ is self-consistently calculated from simulations according to Eq.~(\ref{eq:alpha}) and
$\Sigma_{\rm g}$ is calculated using the higher bound for MMSN.
In the plots, maximum values of the quantities were global maximum throughout the computational grid. 
The central values were averaged across the centre of all the vortices. 
The average value was calculated over the area of all vortices, which gives a lower bound.

\begin{table}
	\centering
	\caption{Summary on the vortex excitation in models listed in Table\,1. Marking "\ding{53}" corresponds to the absent of the given vortex type being multiple small-scale or single large-scale. $\chi_\mathrm{min}$ values show the minimum (i.e., the strongest phase) of vortex aspect ratio for gas/dust components, respectively. $\chi_\mathrm{min}$ values in parenthesis represents the minimum value reached by the end of simulation.  The last column lists the maximum value of total midplane density in terms of Roche density at that radius. Models with asterisk use extra wide $\Delta R_\mathrm {dze}=4H_\mathrm{dze}$ viscosity transition.}
	\label{tbl:RWI}
	\begin{tabular}{lcccccc} 
		\hline
		Mod. & St & \shortstack{Small\\ vort.} & \shortstack{Large \\ vort.} &  \shortstack{ $t_{\rm RWI}$ \\ $(N_{\rm orb})$} & {\bf \shortstack{$\mathbf{\chi_\mathrm{min}}$\\ \rm (dust/gas)}} &
       $\rho_{\rm tot}/\rho_{\rm Roche}$\\
		\hline
		{\bf A1} & $10^{-1}$ &  \ding{53} & \ding{53} & -   & -  & - \\
		{\bf A2} & $10^{-2}$ &  \ding{53} & \ding{52} & 350   & 16/10 & $7.5\times 10^{-2}$\\
		{\bf A3} & $10^{-3}$ &  \ding{53} & \ding{52} & 400   & 16/14 & $9.6\times 10^{-3}$\\
		{\bf A4} & $10^{-4}$ &  \ding{53} & \ding{52} & 400   & 15/15  & $8.6\times 10^{-3}$\\
		{\bf A5} & $10^{-5}$ &  \ding{53} & \ding{52} & 400   & 15/15 & $8.6\times 10^{-3}$ \\
		{\bf A5*} & $10^{-5}$ &  \ding{53} & \ding{52} & 1500 & 34/34  & $7.0\times 10^{-3}$ \\
		\hline
		{\bf B1} & $10^{-1}$ &  \ding{52} & \ding{53} & 50  & 3/7 & 4.9 \\
		{\bf B2} & $10^{-2}$ &  \ding{52} & \ding{53} & 100 & 6/6 & $3.6\times 10^{-2}$\\
		{\bf B2*} & $10^{-2}$ & \ding{52} & \ding{53} & 100   & 5/5 & $3.4\times 10^{-2}$ \\
		{\bf B3} & $10^{-3}$ &  \ding{53} & \ding{52} & 250 & 10/10 & $1.2\times 10^{-2}$\\
		{\bf B4} & $10^{-4}$ &  \ding{53} & \ding{52} & 350 & 14/14 & $9.0\times 10^{-3}$ \\
		{\bf B5} & $10^{-5}$ &  \ding{53} & \ding{52} & 350 & 15/15 & $8.7\times 10^{-3}$ \\
		{\bf B5*} & $10^{-5}$ & \ding{53} & \ding{52} & 950 & 32/32 & $7.2\times 10^{-3}$\\
		\hline
		{\bf C1} & $10^{-1}$ &  \ding{52} & \ding{53} & 50    & 3/7 & 3.3\\
		{\bf C2} & $10^{-2}$ &  \ding{52} & \ding{53} & 150   & 5/6 & $3.2\times 10^{-2}$ \\
		{\bf C3} & $10^{-3}$ &  \ding{53} & \ding{52} & 650   & 9/9 & $7.0\times 10^{-3}$\\
        {\bf C3*}& $10^{-3}$ &  \ding{53}  & \ding{52} & 1350 & 8/8  & $5.1\times 10^{-3}$ \\
		{\bf C4} & $10^{-4}$ &  \ding{53} & \ding{53} & -     & - & - \\
		{\bf C5} & $10^{-5}$ &  \ding{53} & \ding{53} & -     & - & - \\
		\hline
		{\bf D1} & $10^{-1}$ &  \ding{52} & \ding{53} & 50    & 3/10 & 4.4\\
		{\bf D2} & $10^{-2}$ &  \ding{52} & \ding{53} & 100   & 9/9 & $1.2\times 10^{-1}$\\
		{\bf D3} & $10^{-3}$ &  \ding{53} & \ding{52} & 200   & 8/7 & $1.3\times 10^{-2}$\\
		{\bf D4} & $10^{-4}$ &  \ding{53} & \ding{52} & 300   & 13/13 & $9.3\times 10^{-3}$ \\
		{\bf D5} & $10^{-5}$ &  \ding{53} & \ding{52} & 350   & 15/15 & $8.7\times 10^{-3}$ \\
        {\bf D5*}& $10^{-5}$ &  \ding{53}  & \ding{52} & 1400 & 32/32 & $6.8\times 10^{-3}$\\
        \hline
		{\bf E1} & $10^{-1}$ &  \ding{52} & \ding{53} & 50   & 5/9 & 4.4\\
		{\bf E2} & $10^{-2}$ &  \ding{52} & \ding{53} & 100  & 6/7 & $8.8\times 10^{-2}$\\
		{\bf E3} & $10^{-3}$ &  \ding{53} & \ding{52} & 120  & 8/6 & $1.1\times 10^{-2}$\\
		{\bf E4} & $10^{-4}$ &  \ding{53} & \ding{52} & 150  & 10/10 & $1.0\times 10^{-2}$ \\
		{\bf E5} & $10^{-5}$ &  \ding{53} & \ding{52} & 150  & 10/10 & $1.0\times 10^{-3}$ \\
        {\bf E5*}& $10^{-5}$ &  \ding{53}  & \ding{52} & 200 & 10/10 & $9.8\times 10^{-3}$ \\
		\hline
	\end{tabular}
\end{table}

\begin{figure*}
\begin{tabular}{cccc}
 A1 & A2  & A3 & A4 \\
  \includegraphics[width=0.25\textwidth]{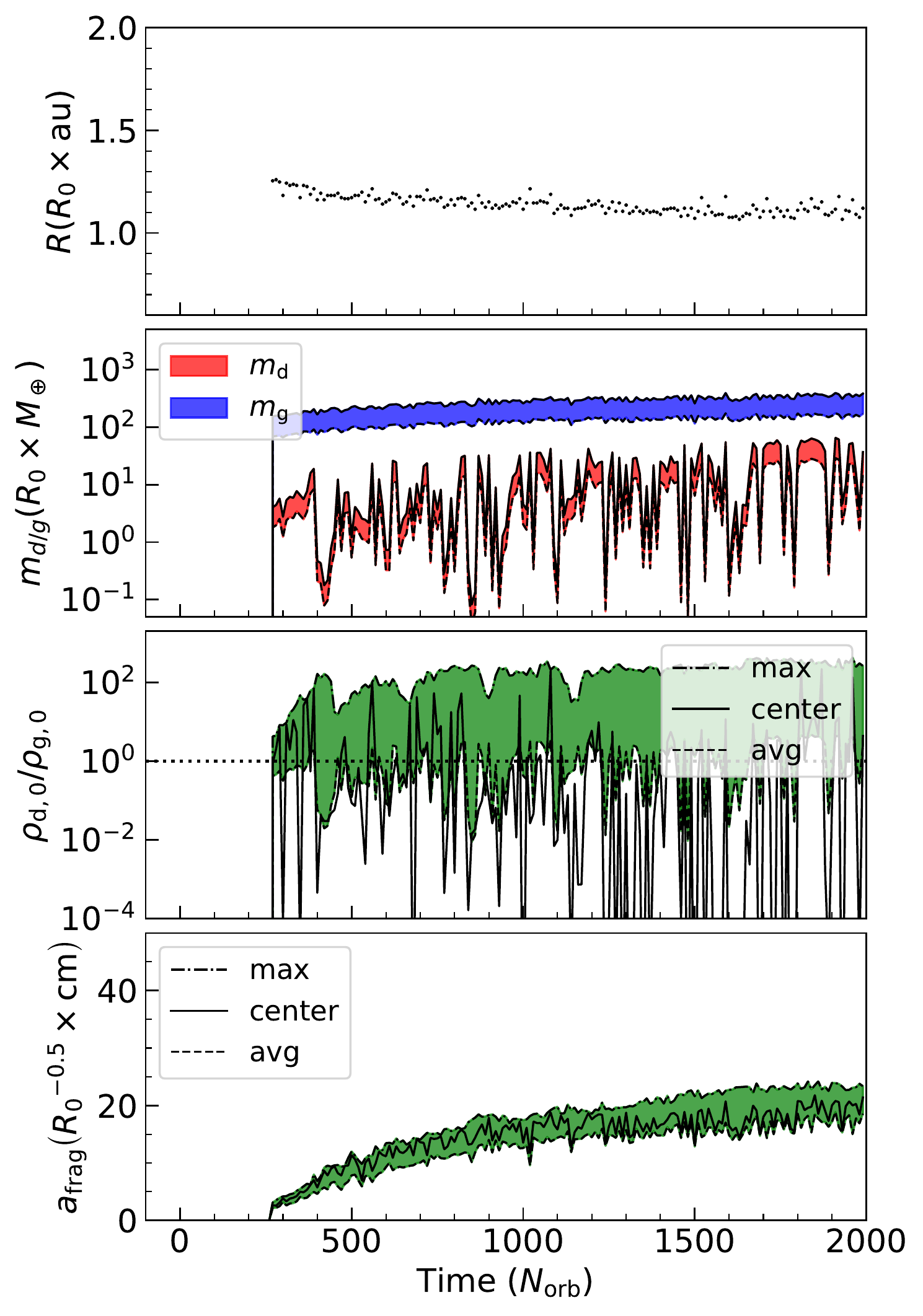} &   \hspace{-0.5cm}\includegraphics[width=0.25\textwidth]{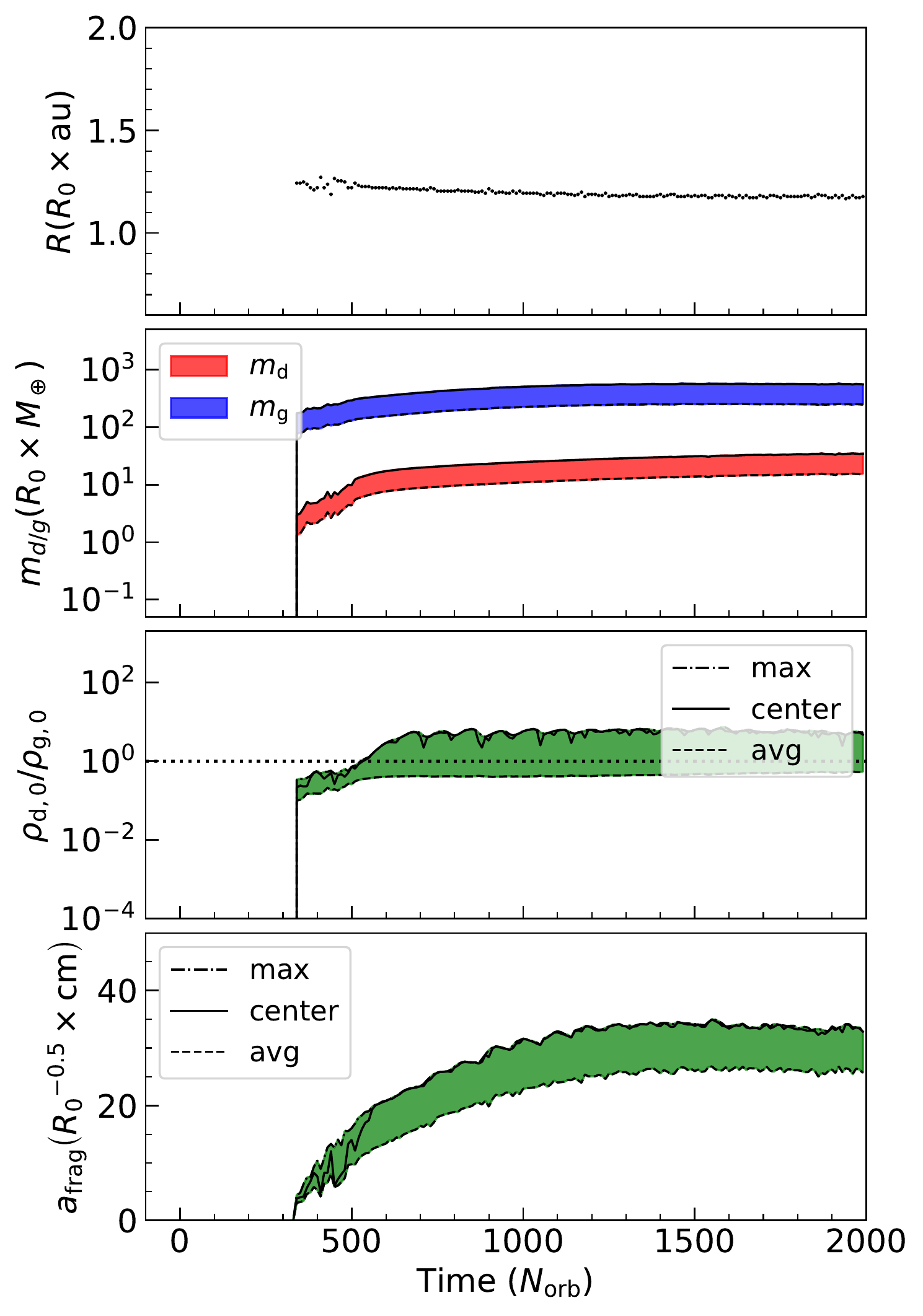} &
  \hspace{-0.5cm}\includegraphics[width=0.25\textwidth]{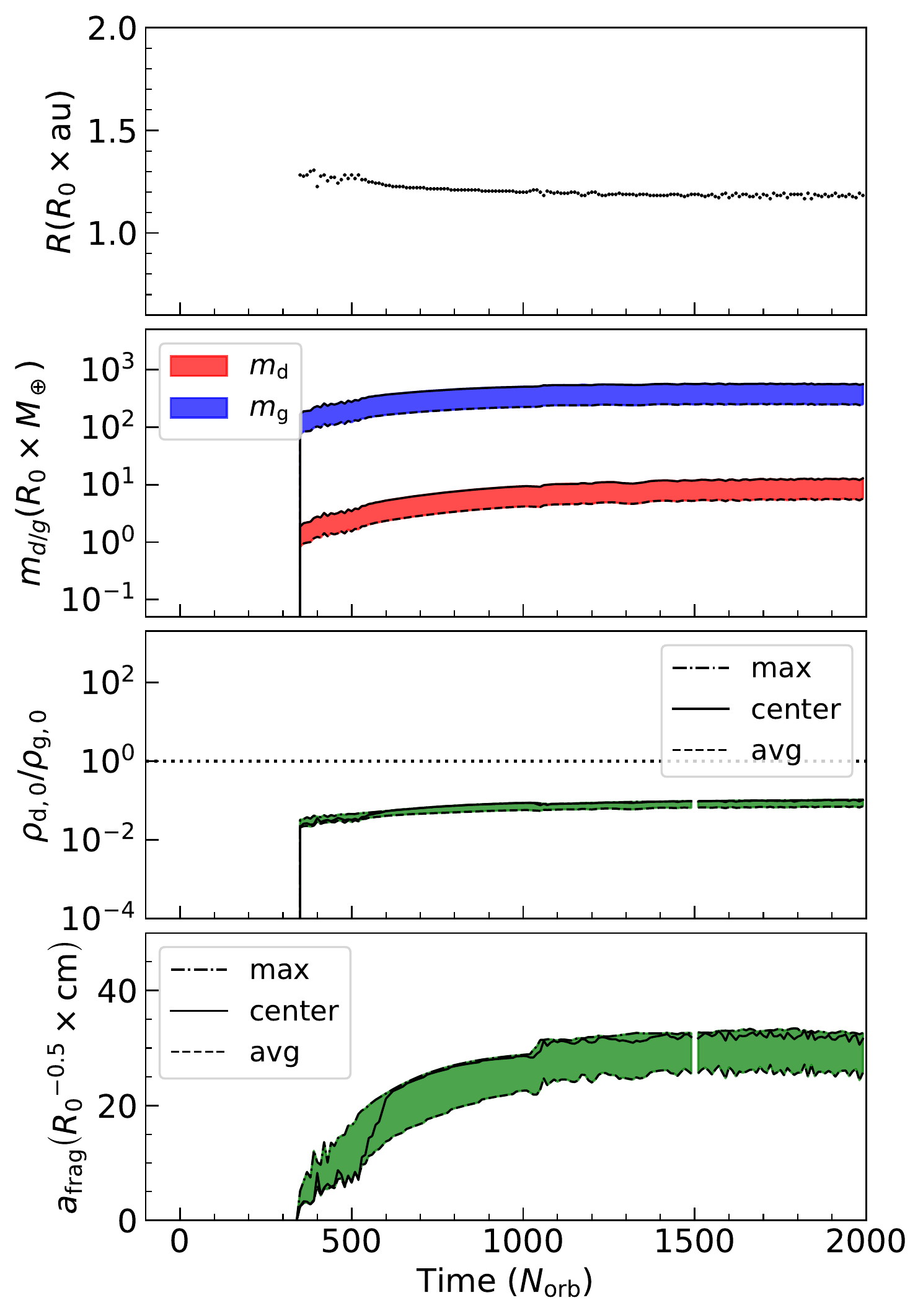} &
  \hspace{-0.5cm}\includegraphics[width=0.25\textwidth]{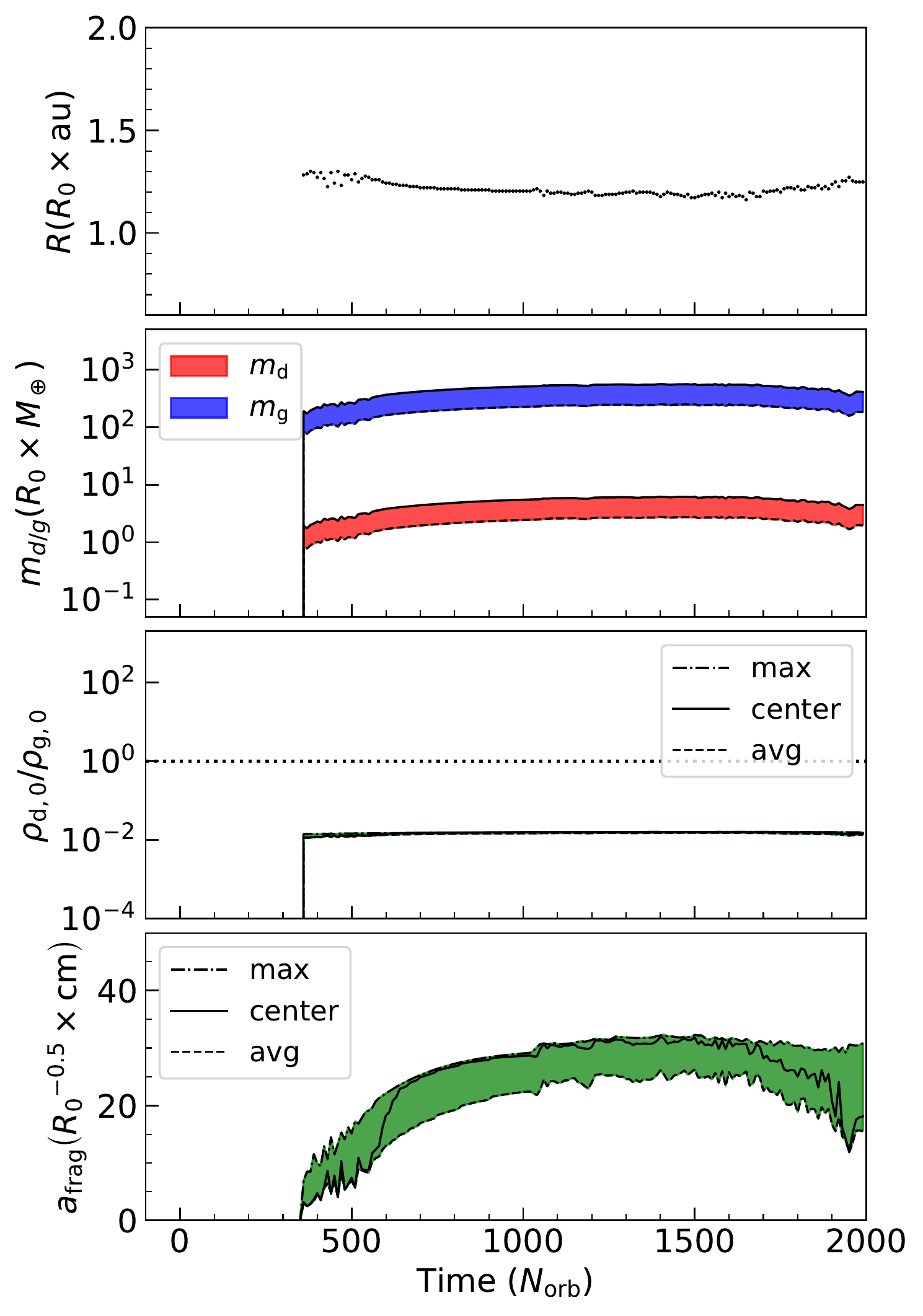}\\
 A5 &  A5*  &  B3 & B4 \\
  \includegraphics[width=0.25\textwidth]{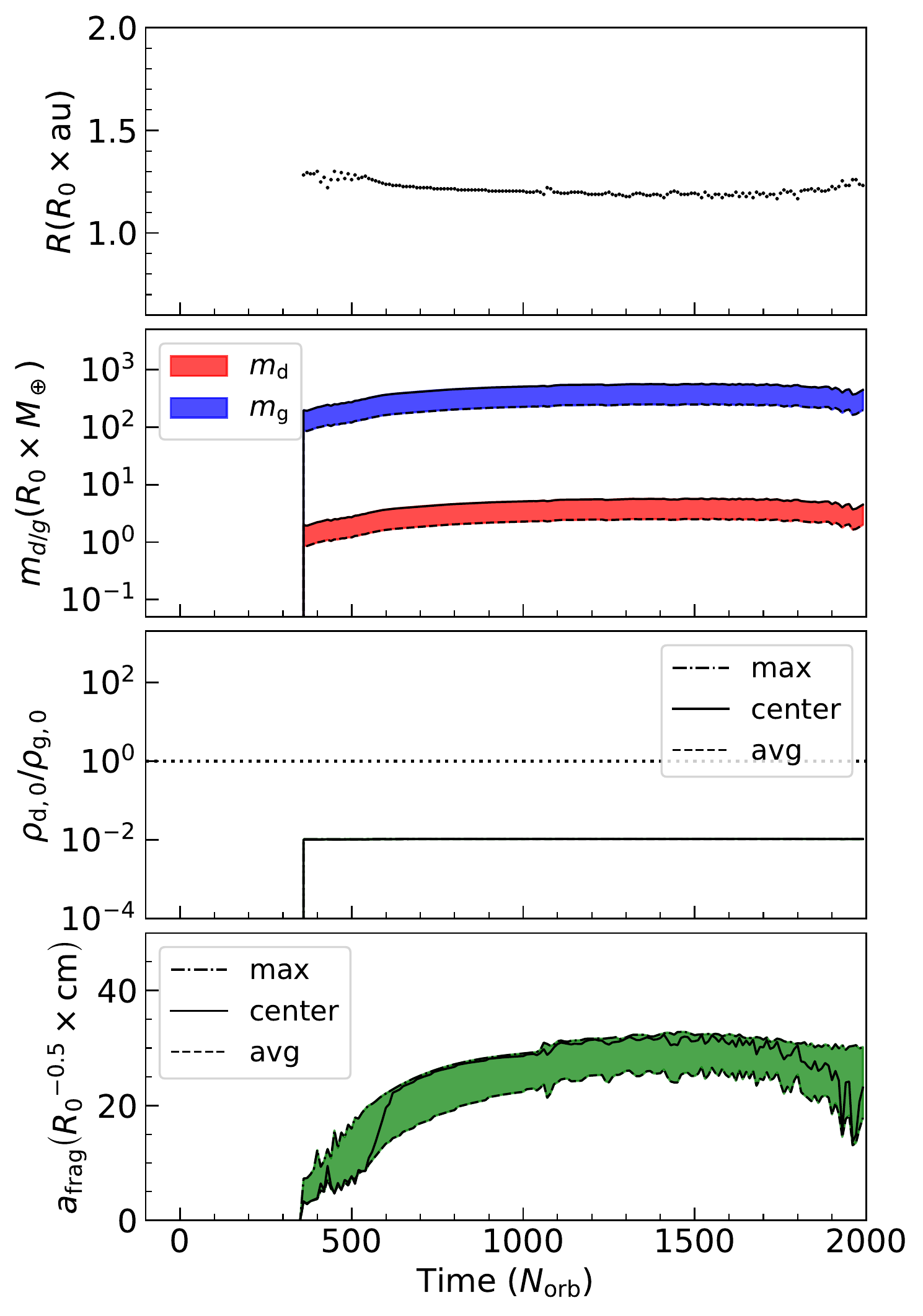} &   \hspace{-0.5cm}\includegraphics[width=0.25\textwidth]{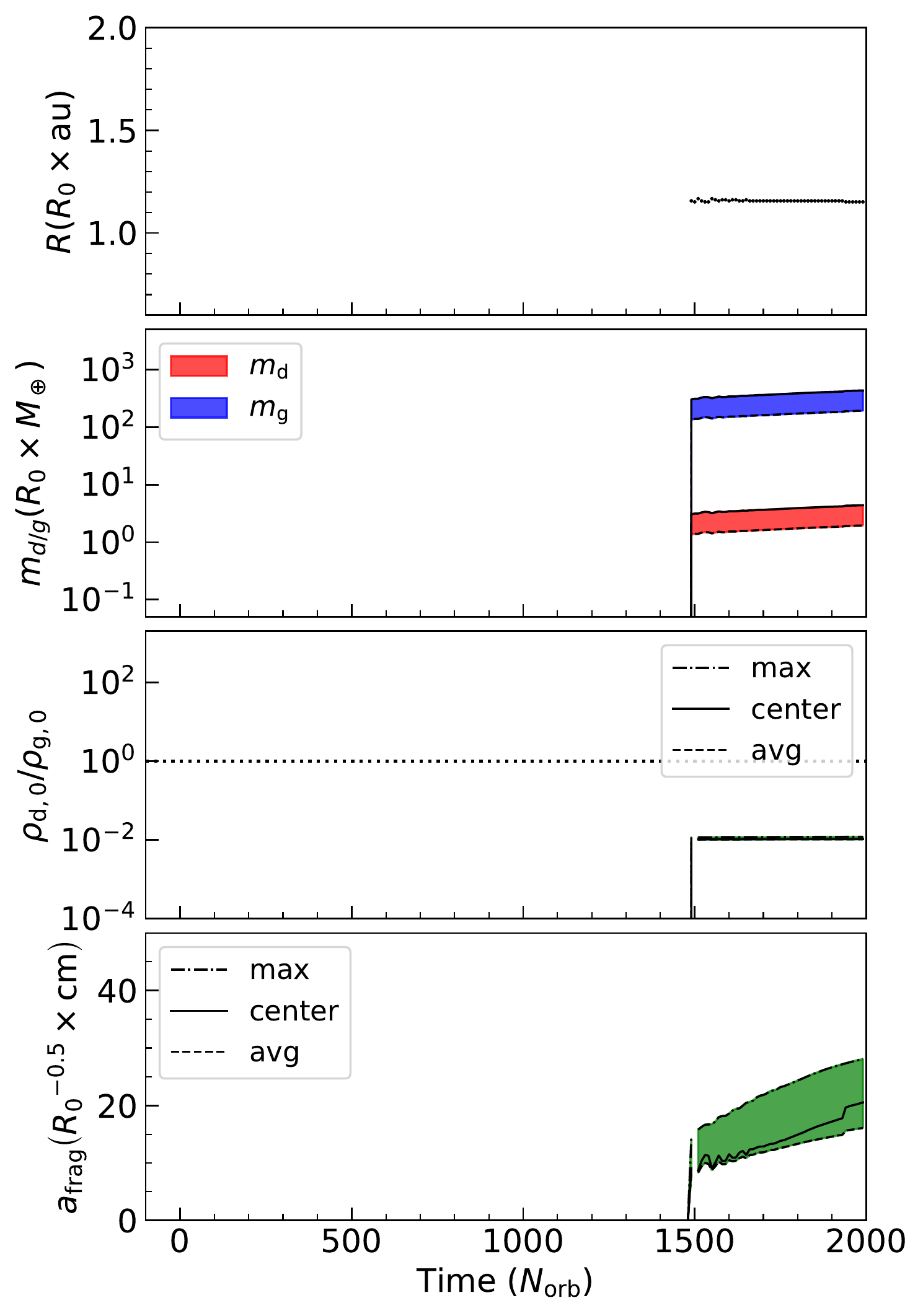} &
  \hspace{-0.5cm}\includegraphics[width=0.25\textwidth]{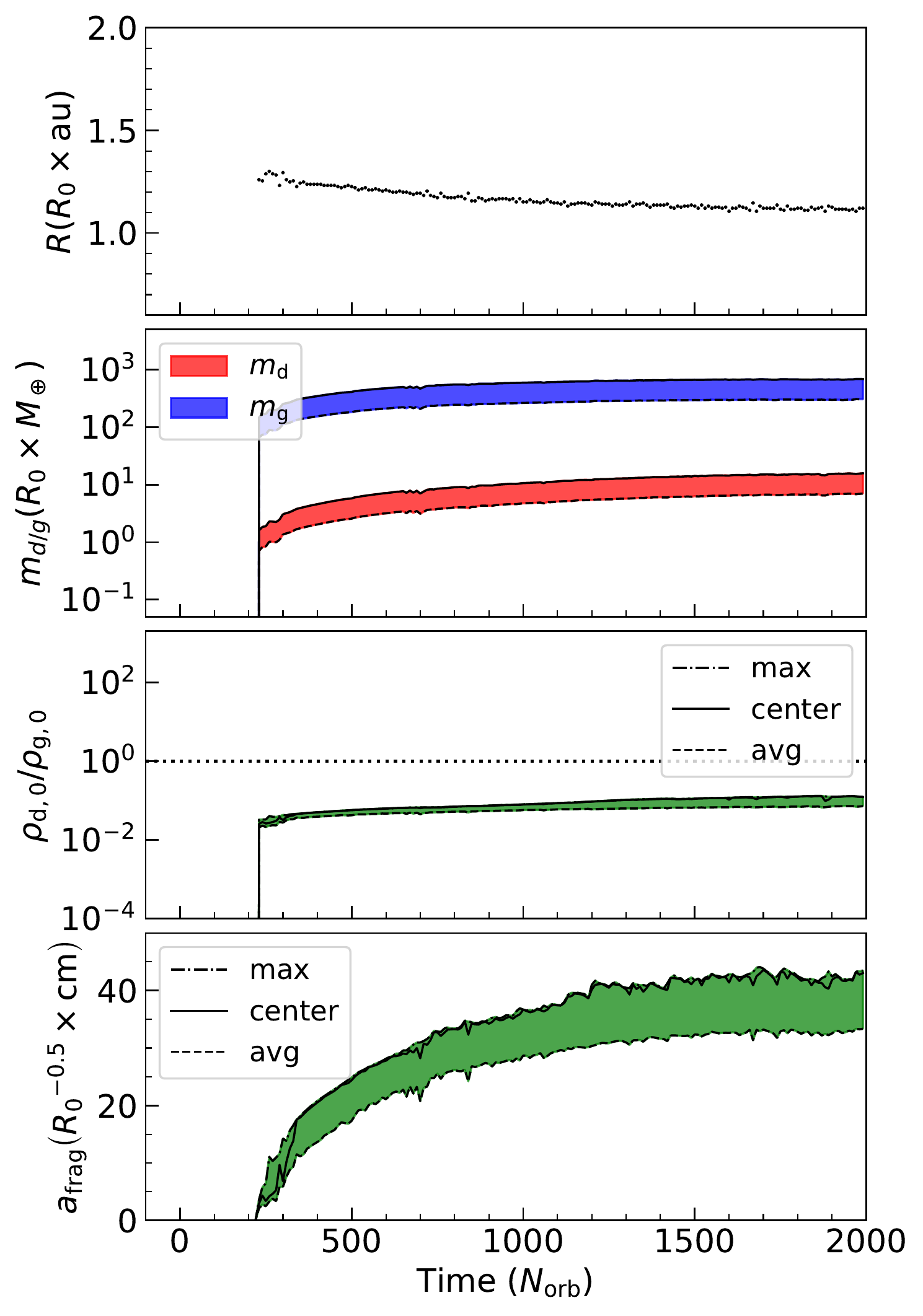} & 
  \hspace{-0.5cm}\includegraphics[width=0.25\textwidth]{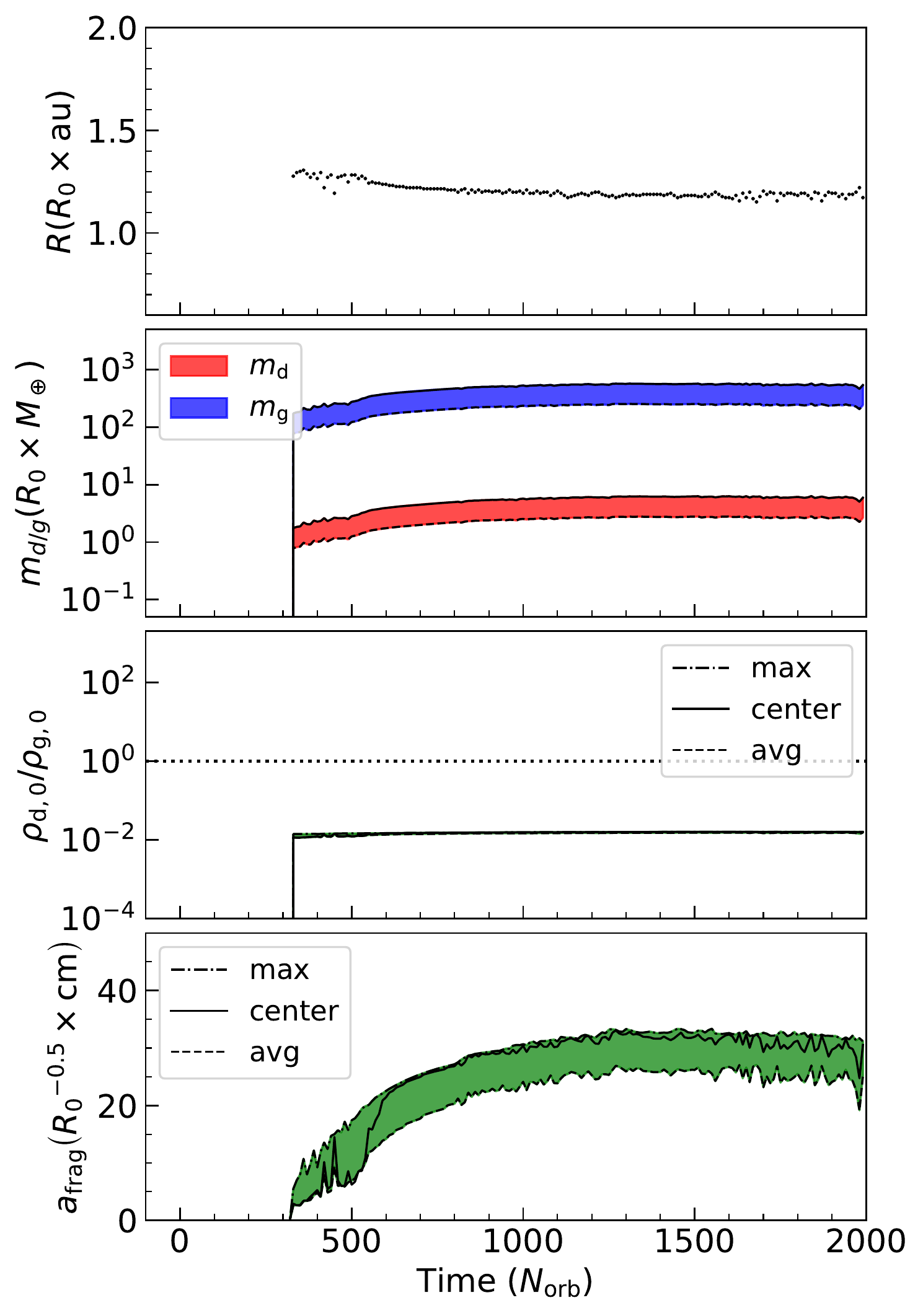} \\
 B5 &  B5*  &  C3 &  C3* \\
  \includegraphics[width=0.25\textwidth]{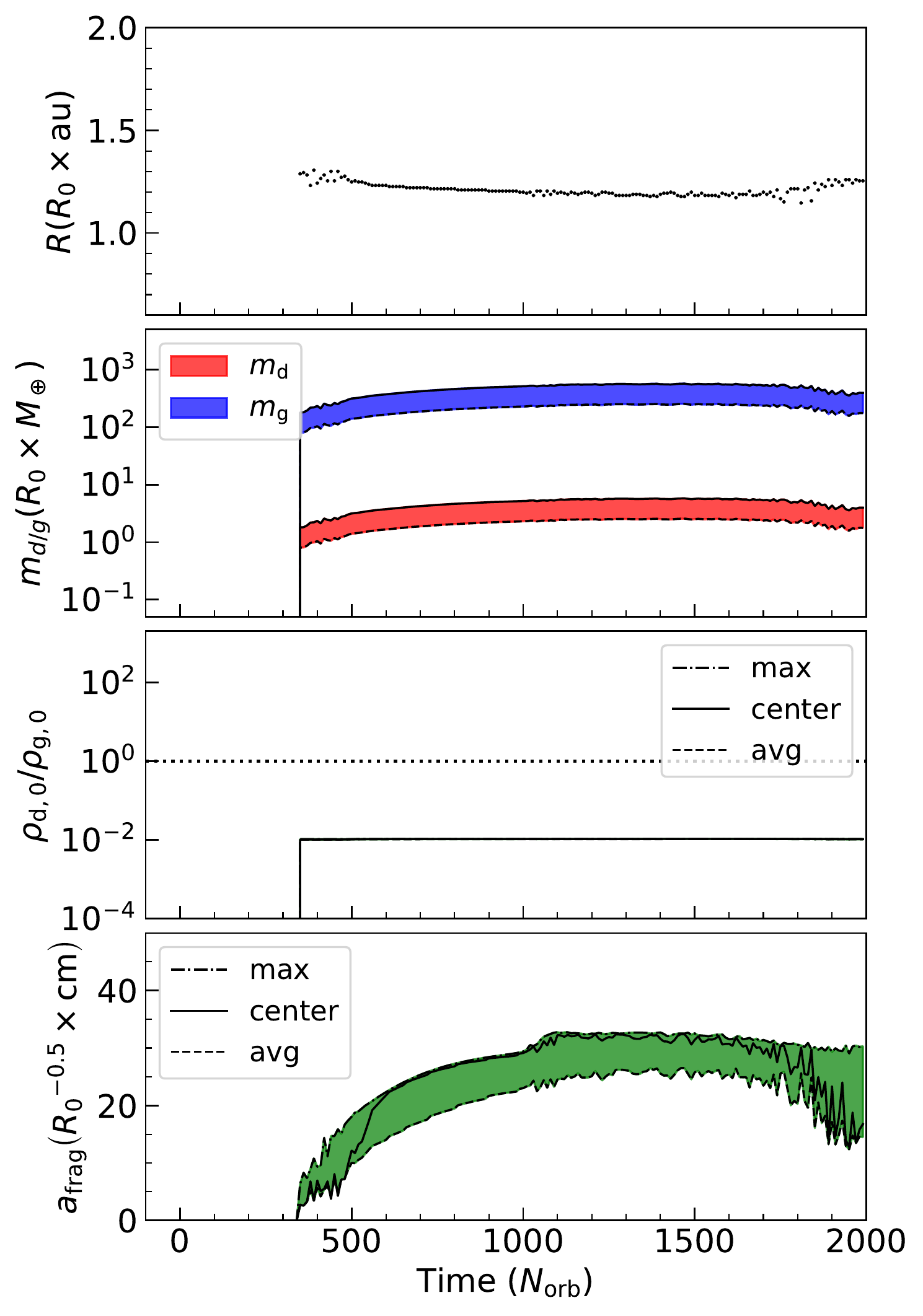} &   \hspace{-0.5cm}\includegraphics[width=0.25\textwidth]{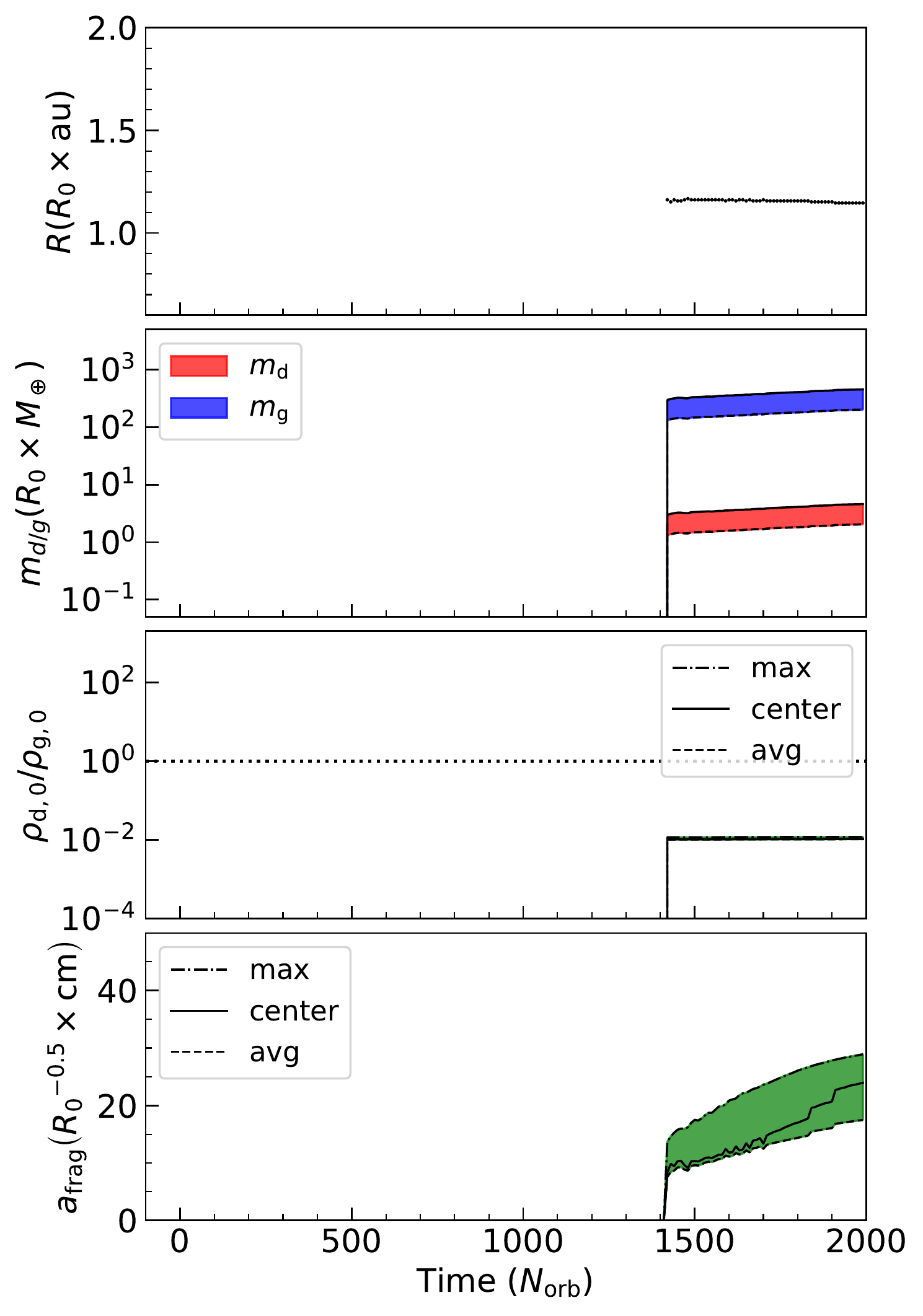} &
  \hspace{-0.5cm}\includegraphics[width=0.25\textwidth]{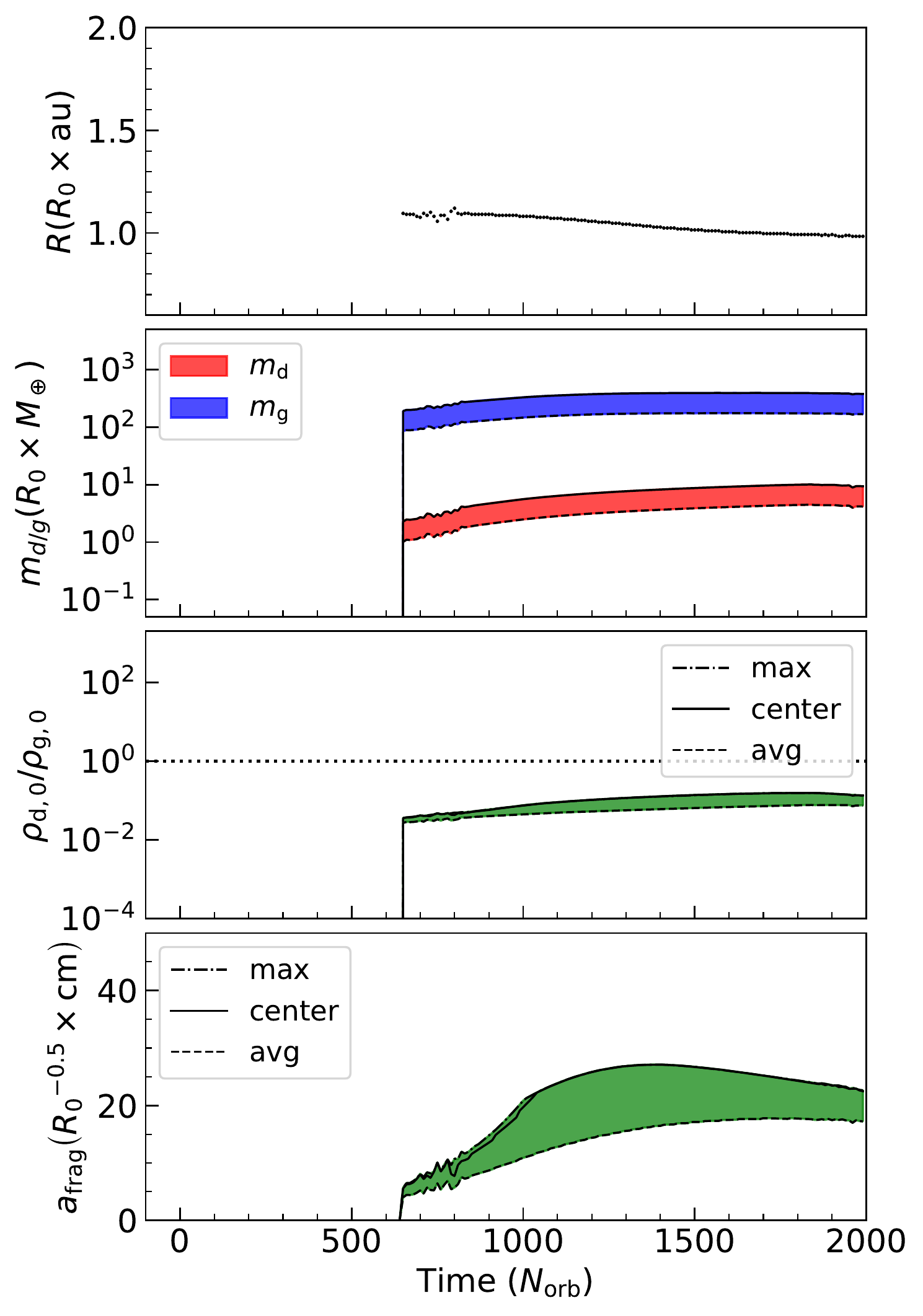} & 
  \hspace{-0.5cm}\includegraphics[width=0.25\textwidth]{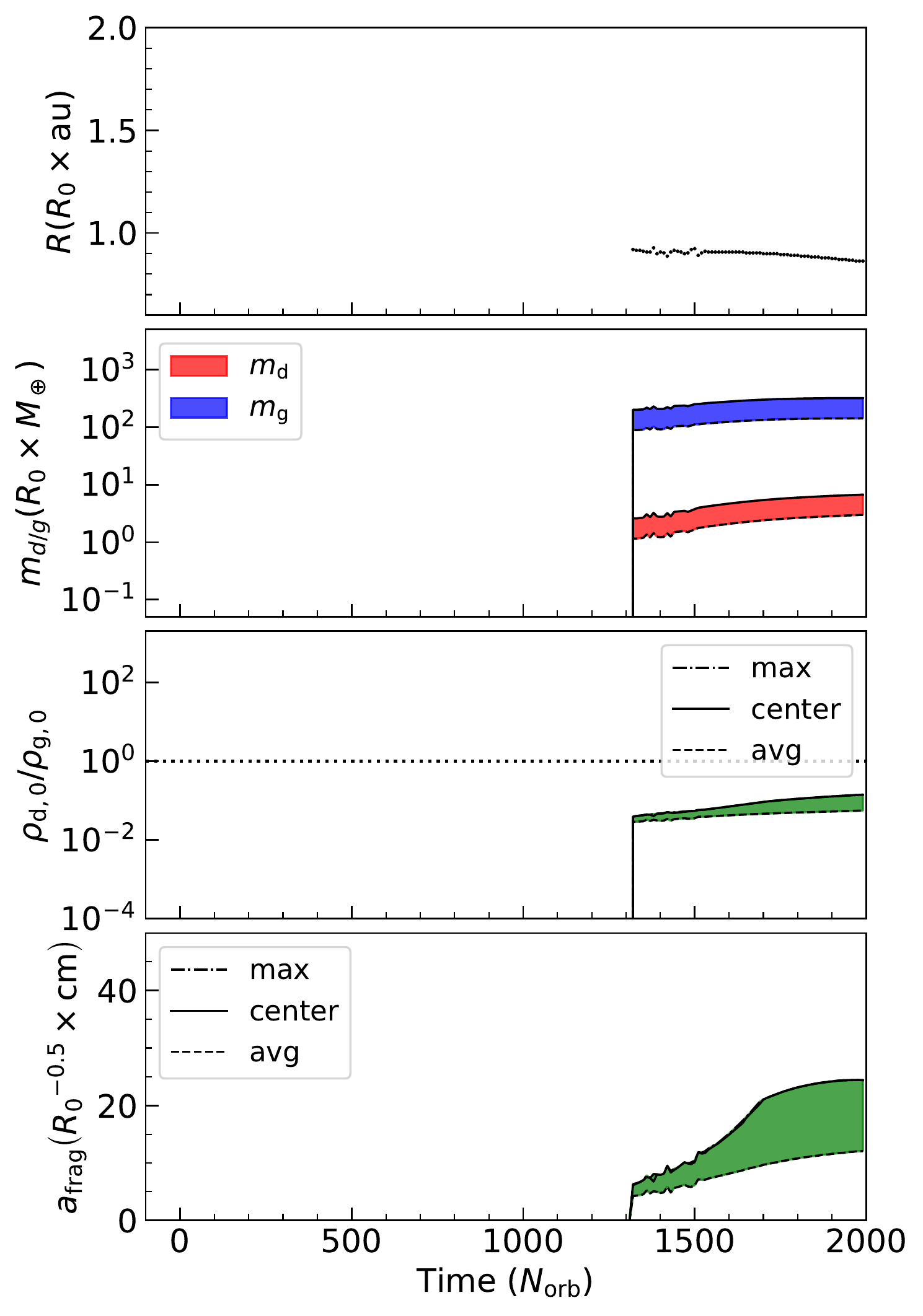} \\  
\end{tabular}
\caption{Analysis on vortex evolution for models A1, A2, A3, A4, A5, A5*, B3, B4, B5, B5*, C3, where C3*, where only a single large-scale vortex develops. Four panels are shown for each models. Panels are from top to bottom: radial distance of the vortex centre, dust (red) and gas (blue) mass inside the vortex, dust-to-gas mass ratio (green) at the vortex centre, fragmentation size, i.e., the maximum size of solid constituent that can be reached due to dust growth process inside the vortex. The shaded regions in gas–dust mass panels correspond to the maximum/minimum values calculated using the upper/lower bounds of gas density for the MMSN model. The dotted lines in panels~e and f show the excitation criterion for triggering streaming instability and the metre size, respectively. Two distinct groups can be identified based on the excitation time of RWI: excitation occurs before and after $N_\mathrm{orb}=500$ in the early and late onset group, respectively.}
\label{fig:single–analysis-1}
\end{figure*}

\begin{figure*}
\begin{tabular}{cccc}
 D3 & D4  & D5 & D5* \\
  \includegraphics[width=0.25\textwidth]{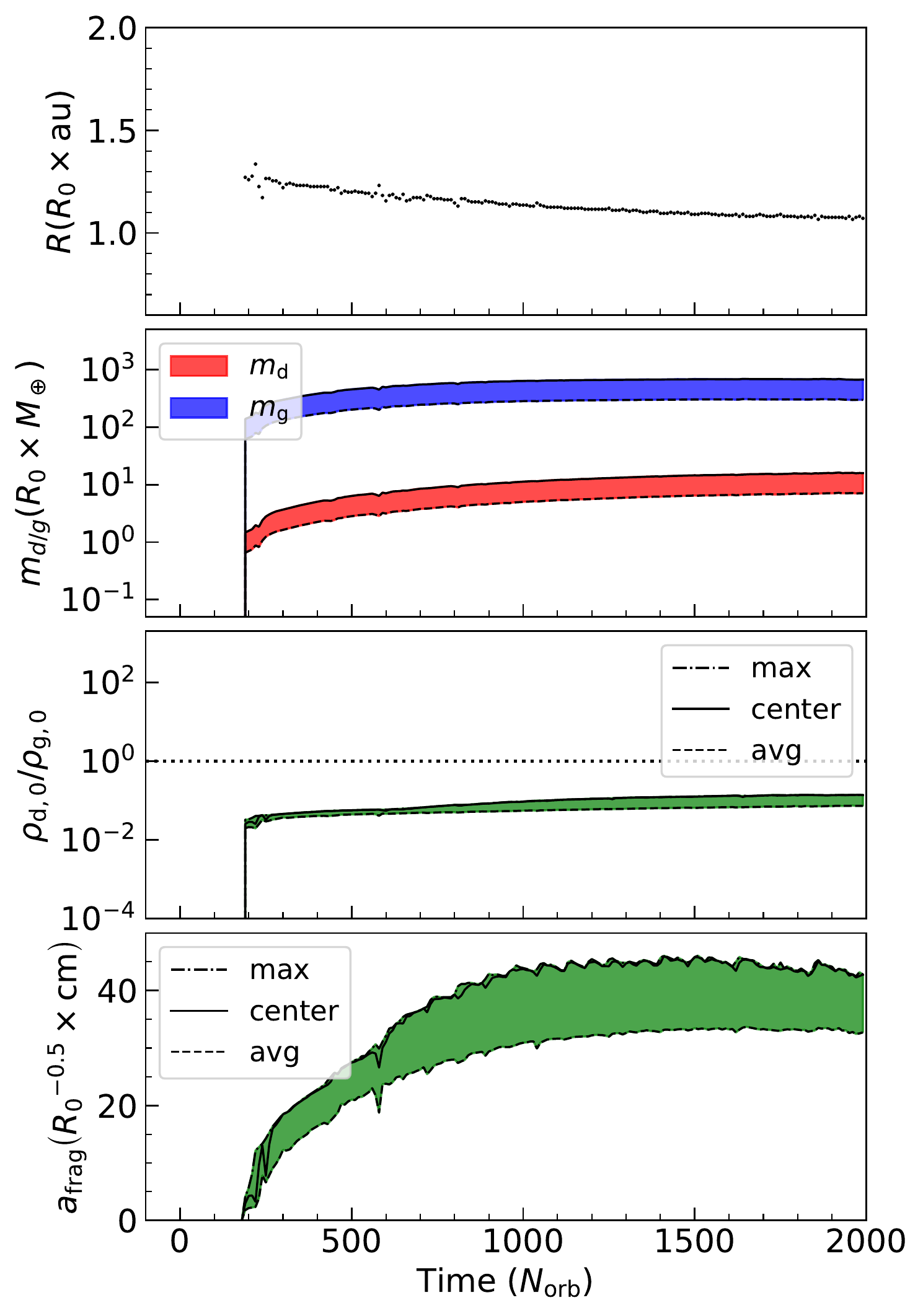} &   \hspace{-0.5cm}\includegraphics[width=0.25\textwidth]{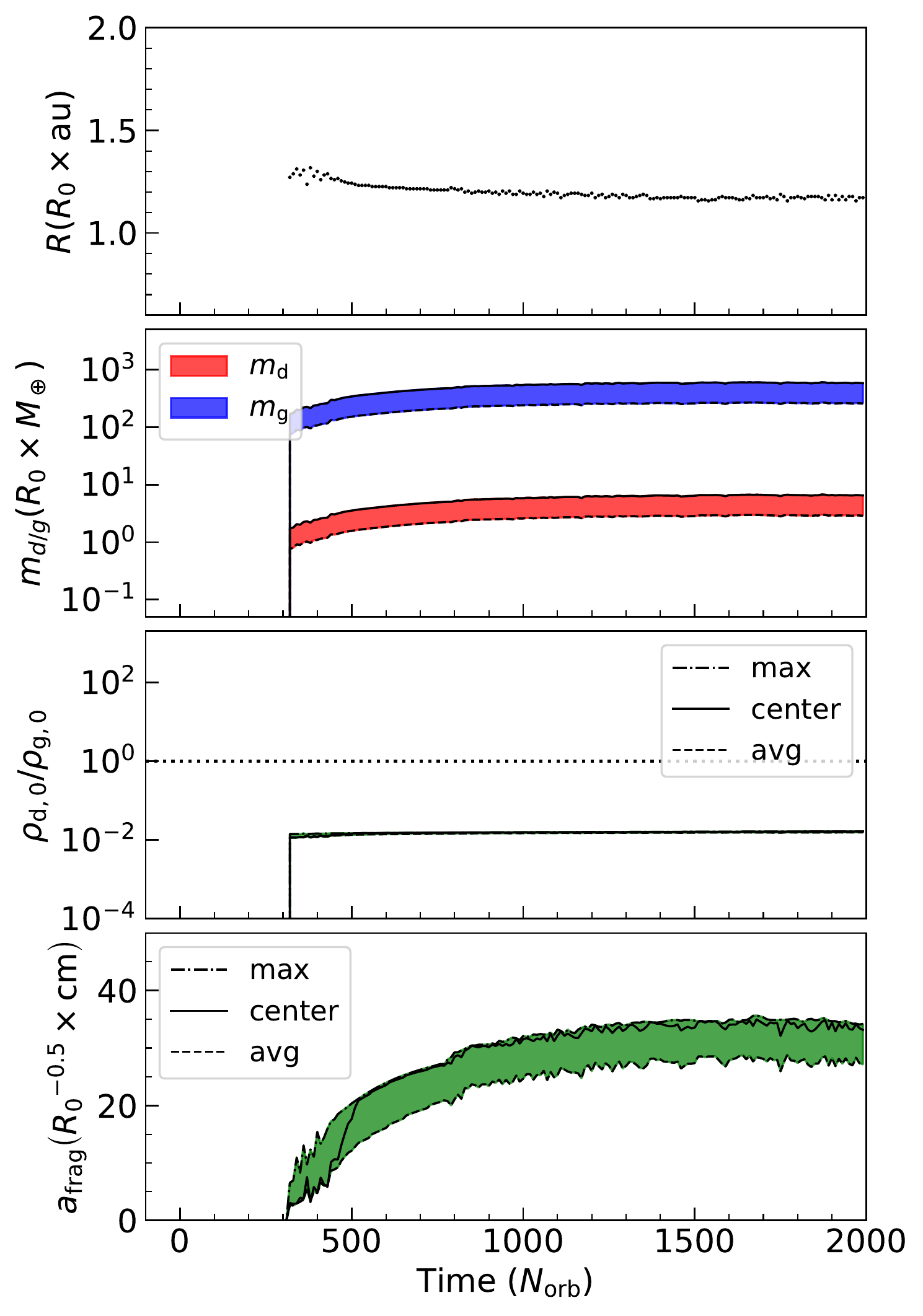} &
  \hspace{-0.5cm}\includegraphics[width=0.25\textwidth]{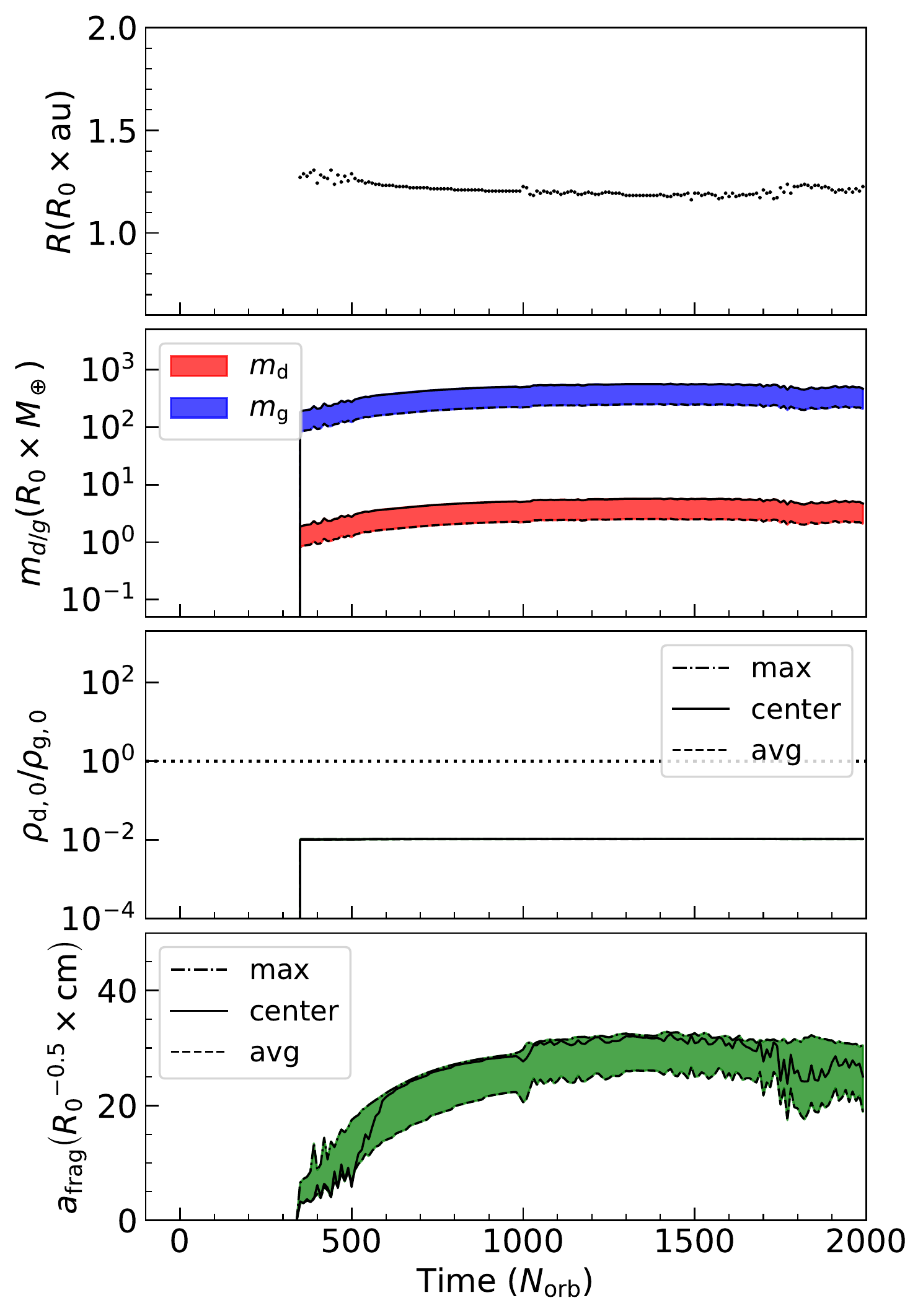} &
  \hspace{-0.5cm}\includegraphics[width=0.25\textwidth]{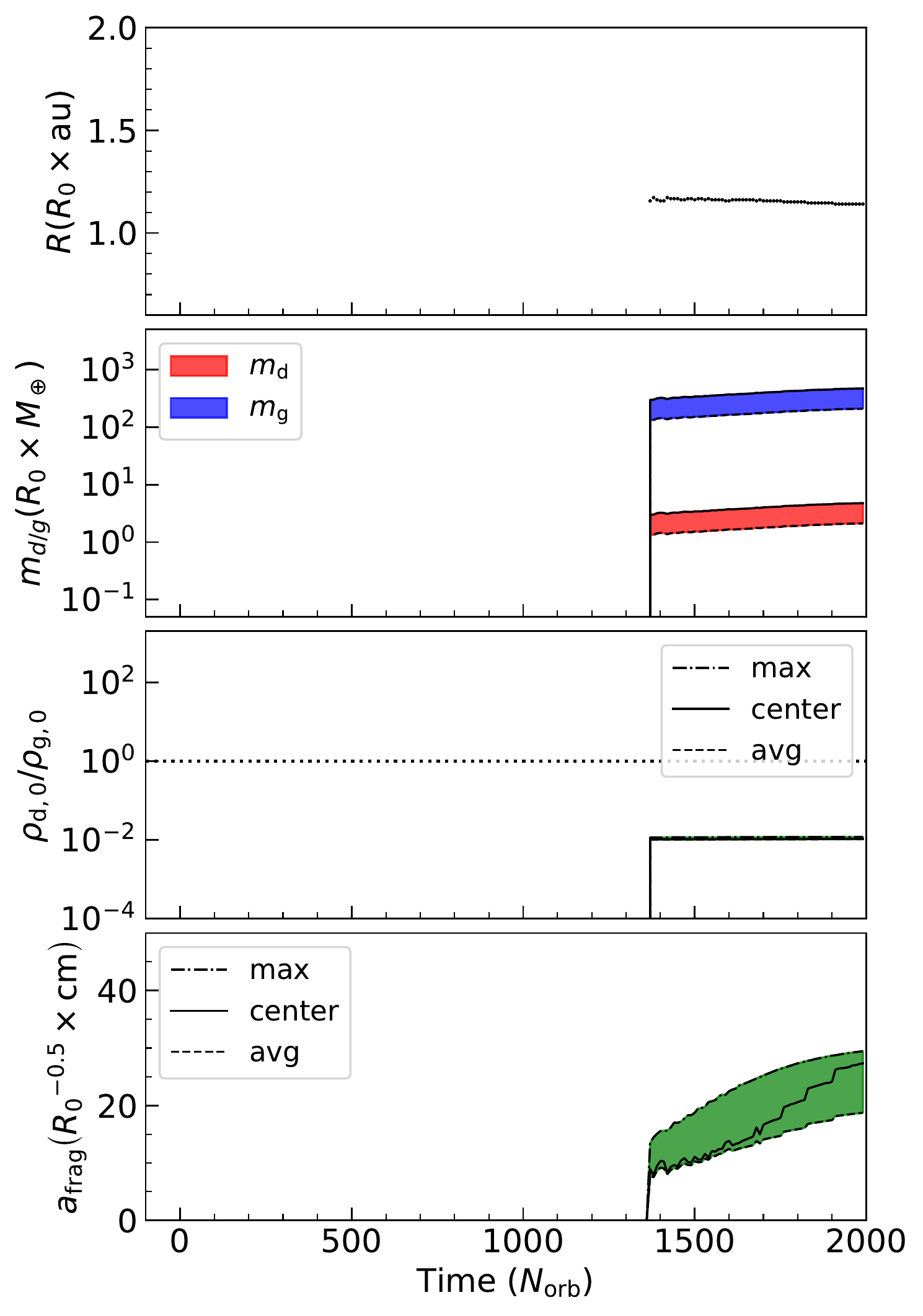}\\
 E3 & E4  & E5 & E5* \\
  \includegraphics[width=0.25\textwidth]{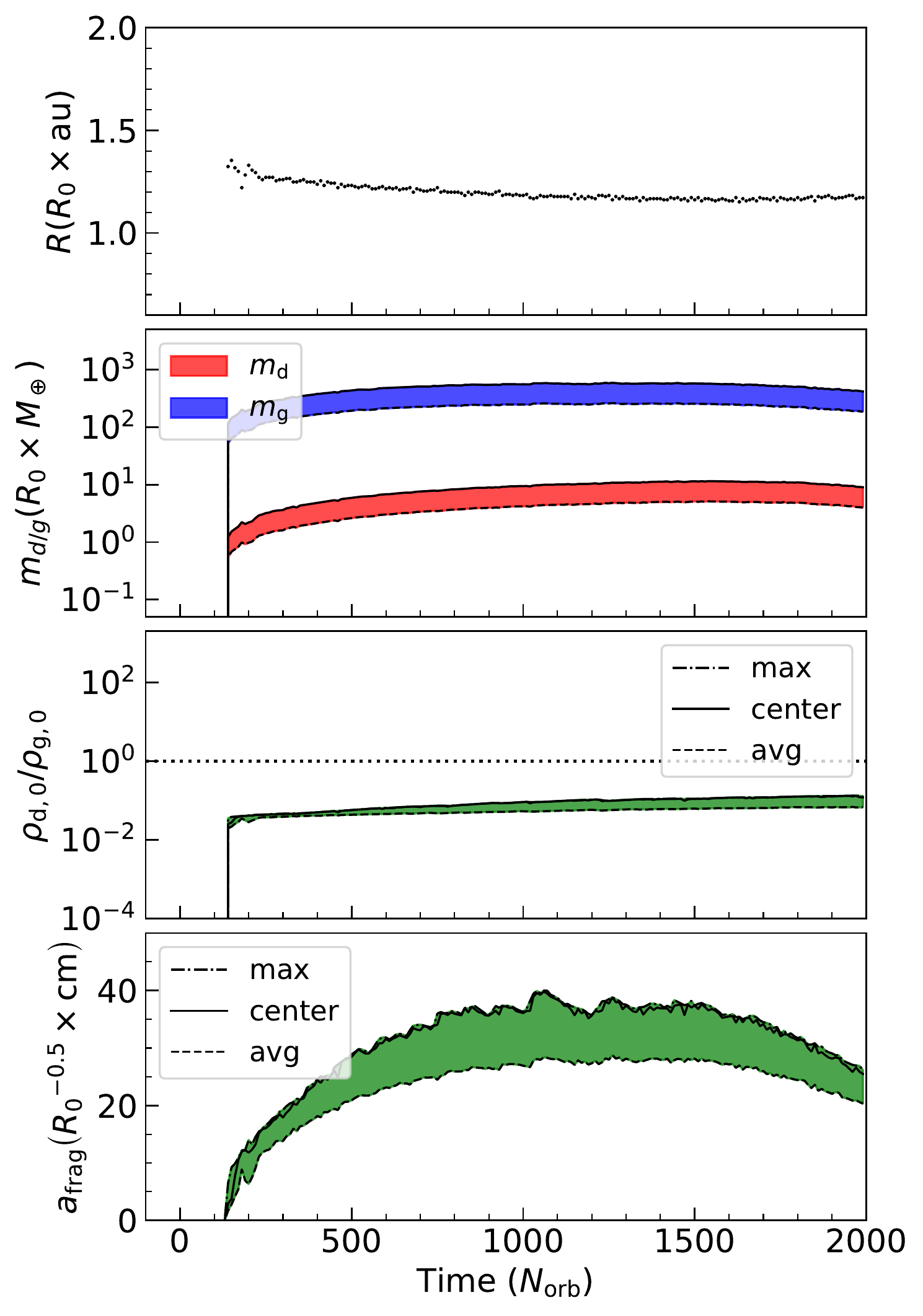} &   \hspace{-0.5cm}\includegraphics[width=0.25\textwidth]{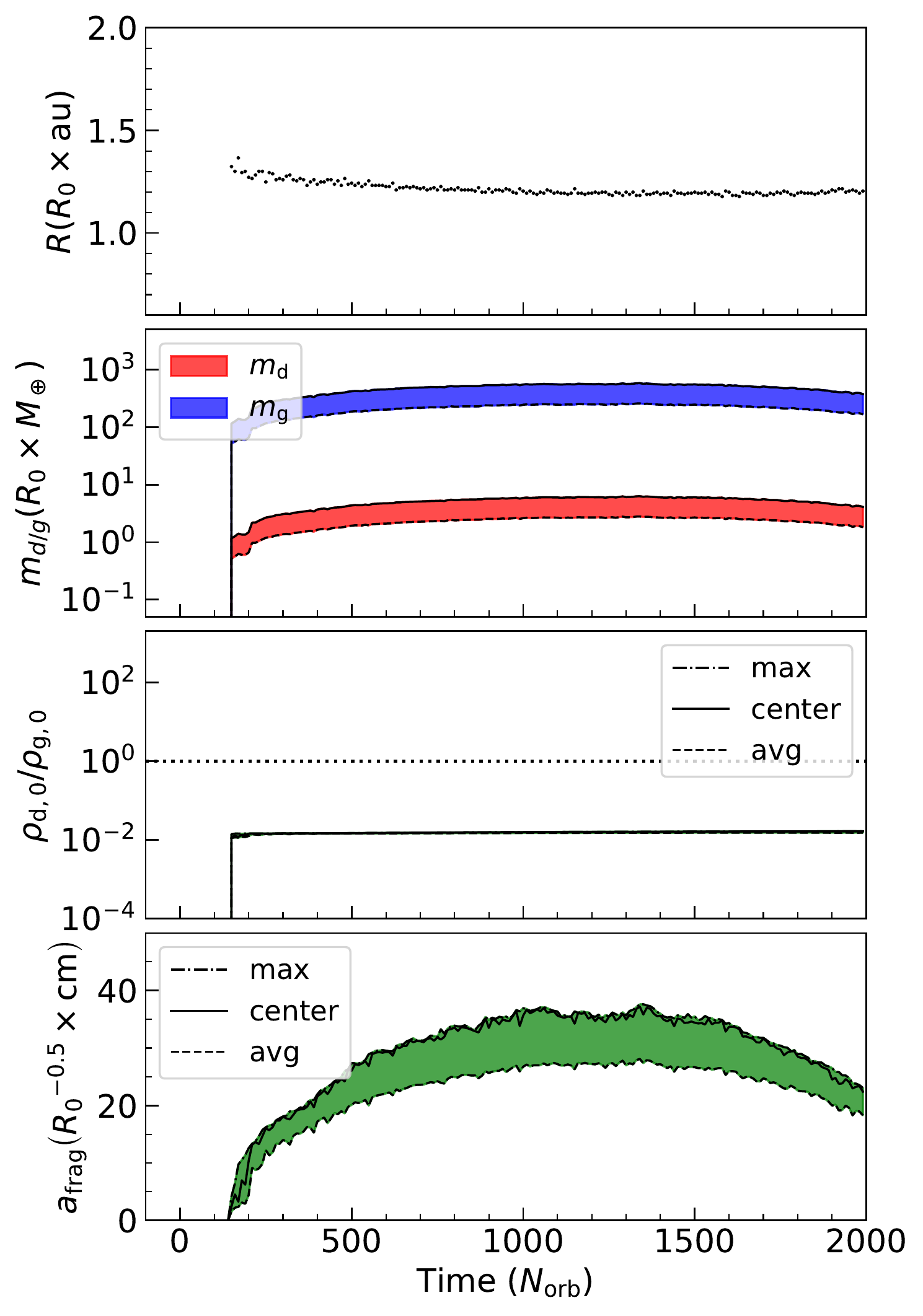} &
  \hspace{-0.5cm}\includegraphics[width=0.25\textwidth]{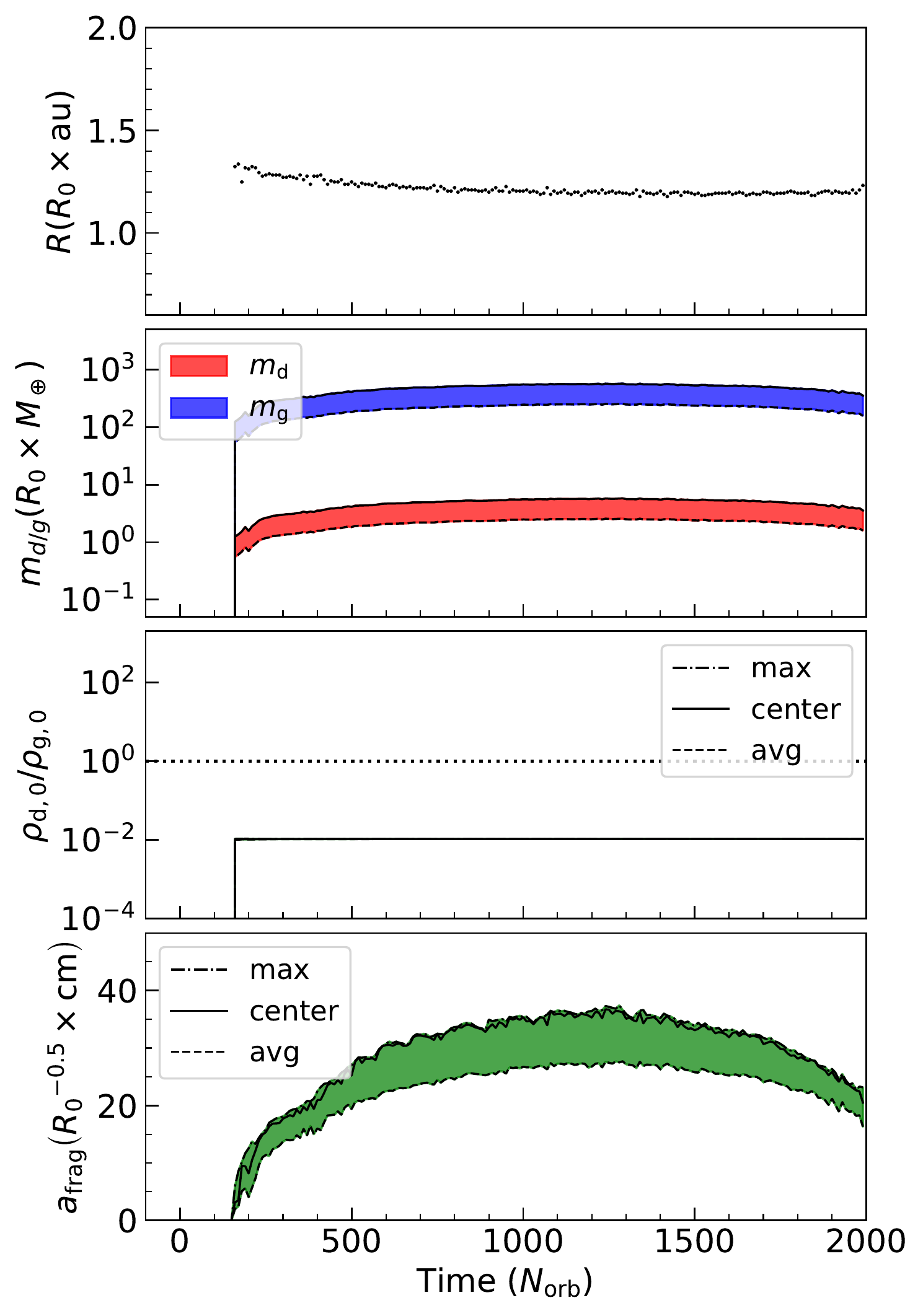} & 
  \hspace{-0.5cm}\includegraphics[width=0.25\textwidth]{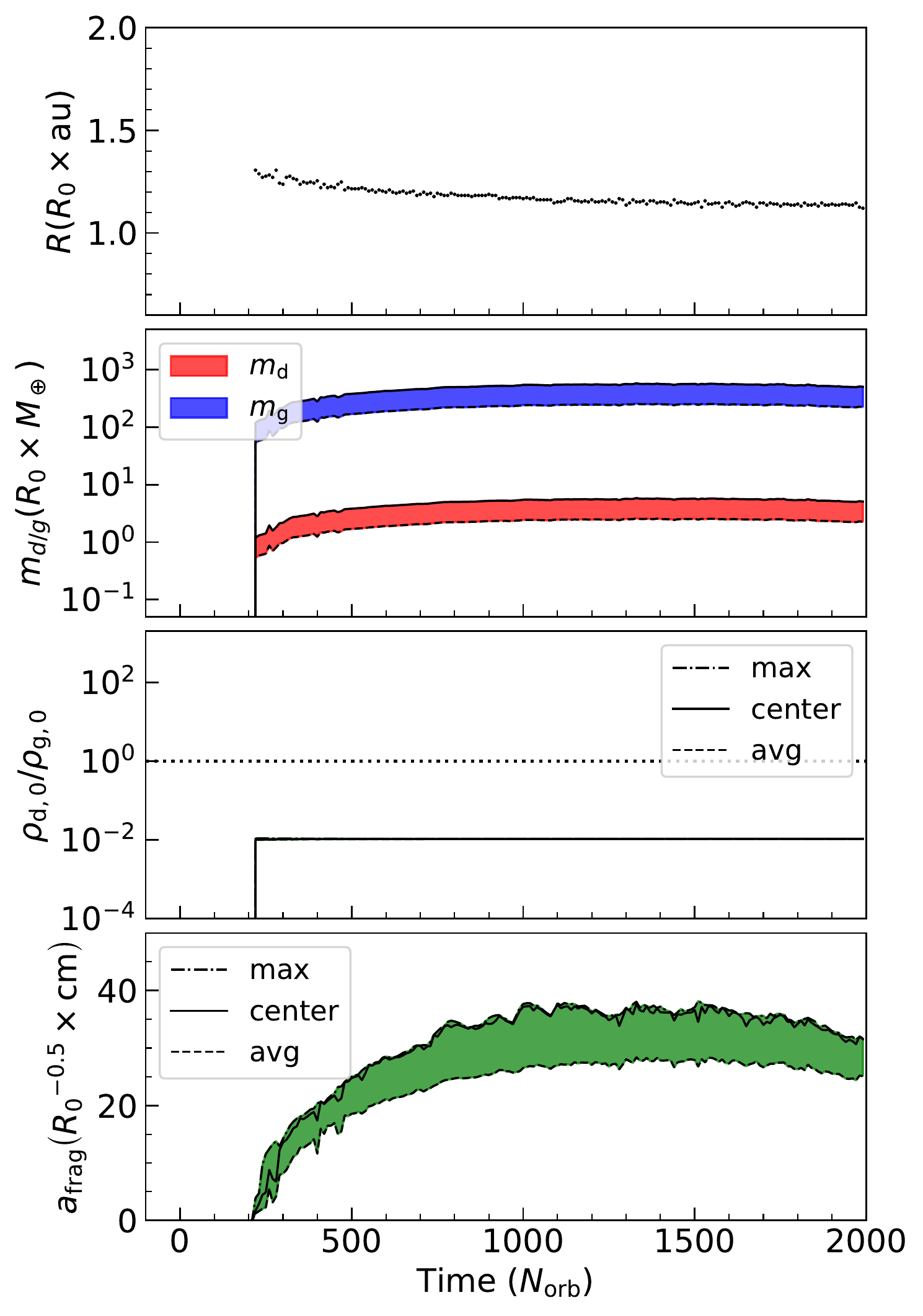} \\
\end{tabular}
\caption{Same as Fig.\,\ref{fig:single–analysis-2}, but for models D3, D4, D5, D5*, E3, E4, E5, E5*, where also a single large-scale vortex develops.}
\label{fig:single–analysis-2}
\end{figure*}

\begin{figure*}
\begin{tabular}{cccc}
 B1 &  C1  &  D1 & E1 \\
  \includegraphics[width=0.25\textwidth]{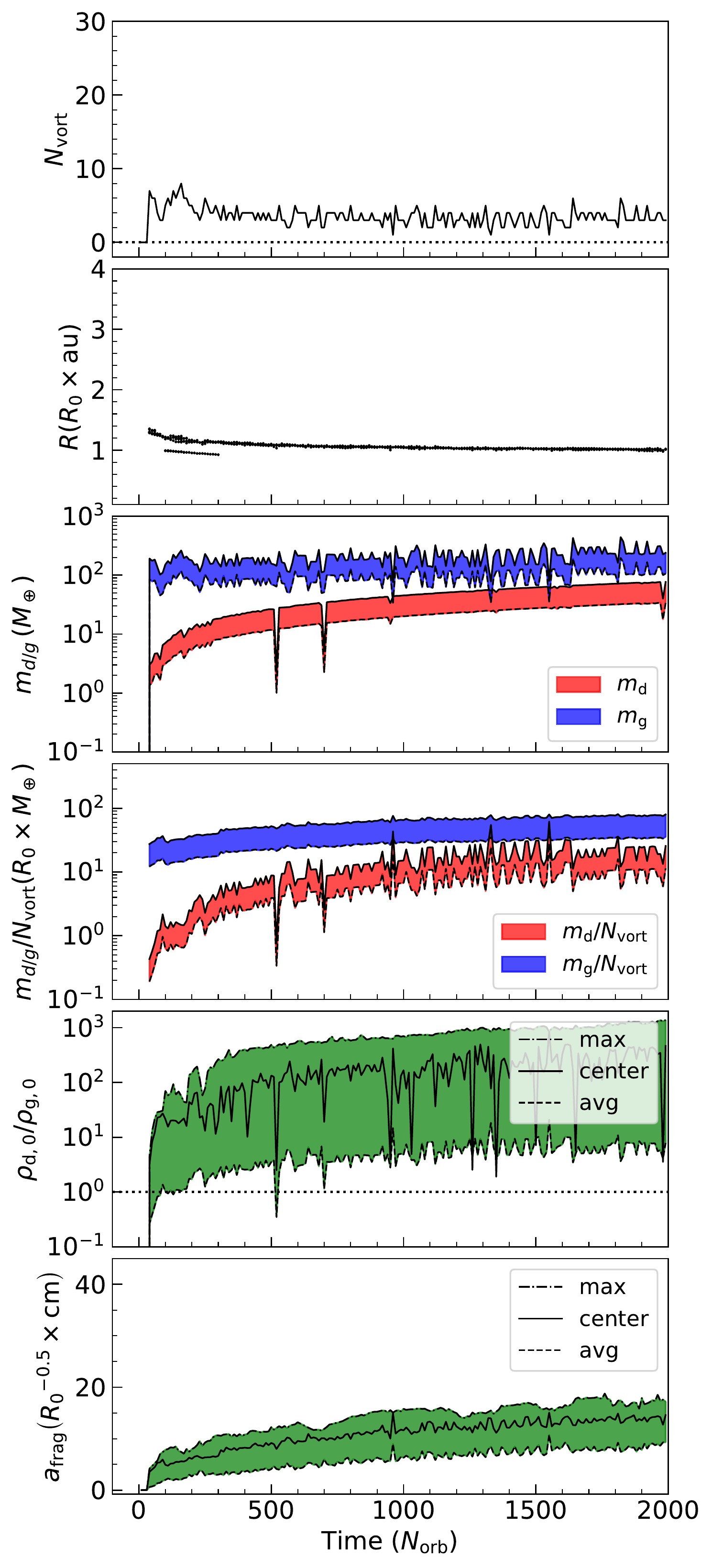} &   \hspace{-0.5cm}\includegraphics[width=0.25\textwidth]{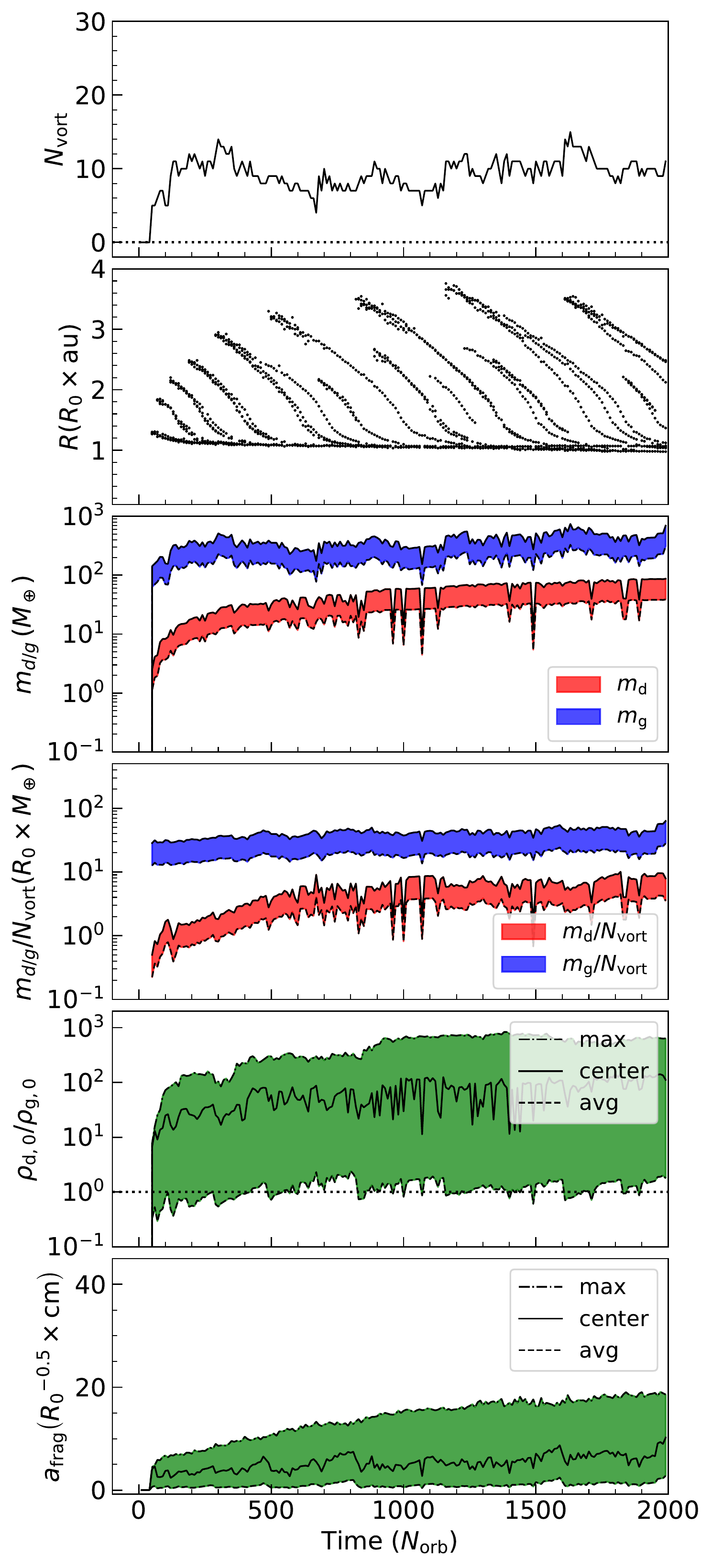} &
  \hspace{-0.5cm}\includegraphics[width=0.25\textwidth]{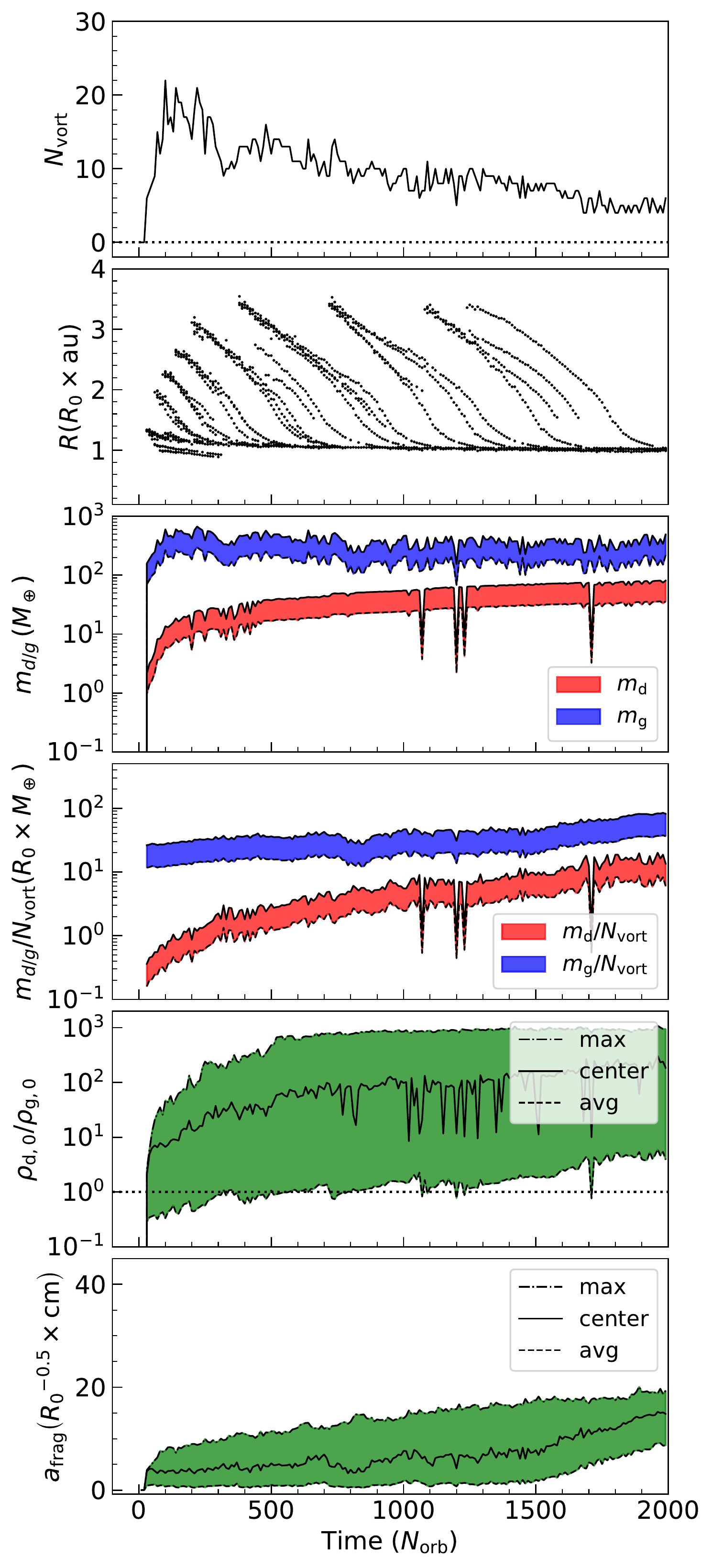} &
  \hspace{-0.5cm}\includegraphics[width=0.25\textwidth]{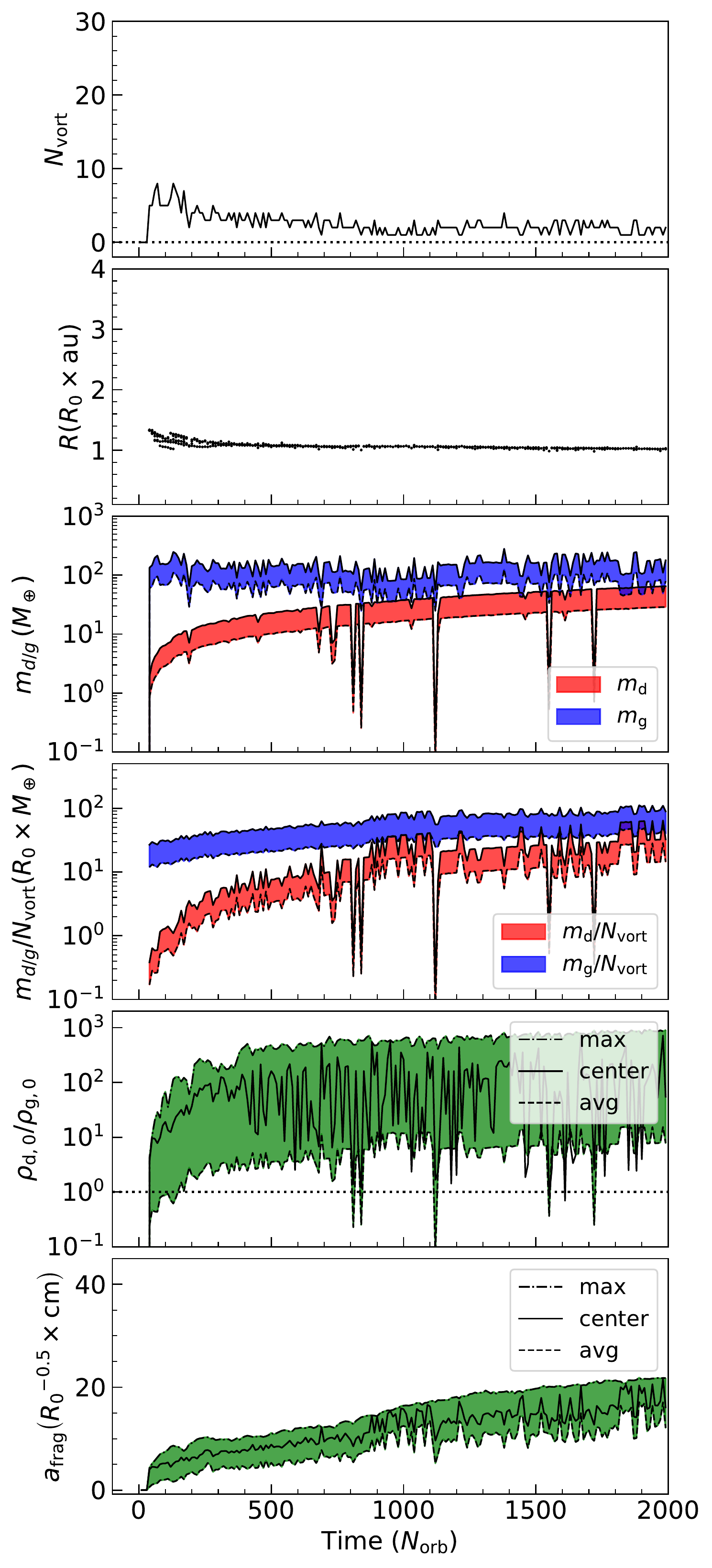}\\
\end{tabular}
\caption{Analysis on vortex evolution for models B1, C1, D1, and E1, where multiple vortices develop. Six panel are shown for each models. Panels are from top to bottom: number of vortices in the disc, radial distances of vortex centres, dust (red) and gas (blue) mass inside the largest vortex, vortex dust (red) and gas (blue) mass content averaged through all vortices, dust-to-gas mass ratio (green) at the vortex centre, fragmentation size, i.e. the maximum size of solid constituent that can be reached due to dust growth process inside the vortex. The shaded regions in gas–dust mass panels correspond to the maximum/minimum values calculated using the upper/lower bounds of gas density for the MMSN model. In models C1 and D1, vortex generation are not concentrated to the viscosity transition ($R=1$), rather shows a cascade like feature.}
\label{fig:multi–analysis-1}
\end{figure*}

\begin{figure*}
\begin{tabular}{cccc}
 B2 & C2  & D2 & E2 \\
  \includegraphics[width=0.25\textwidth]{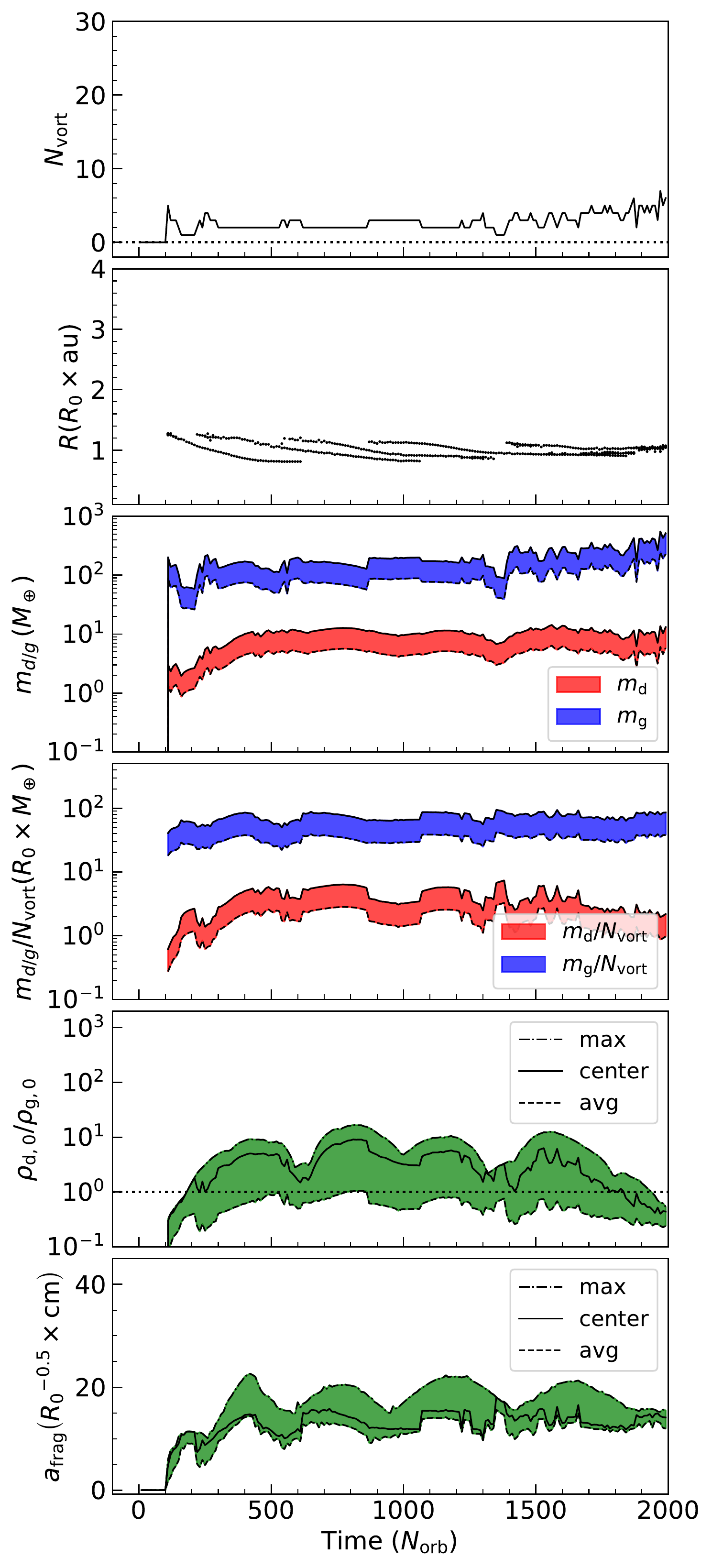} &   \hspace{-0.5cm}\includegraphics[width=0.25\textwidth]{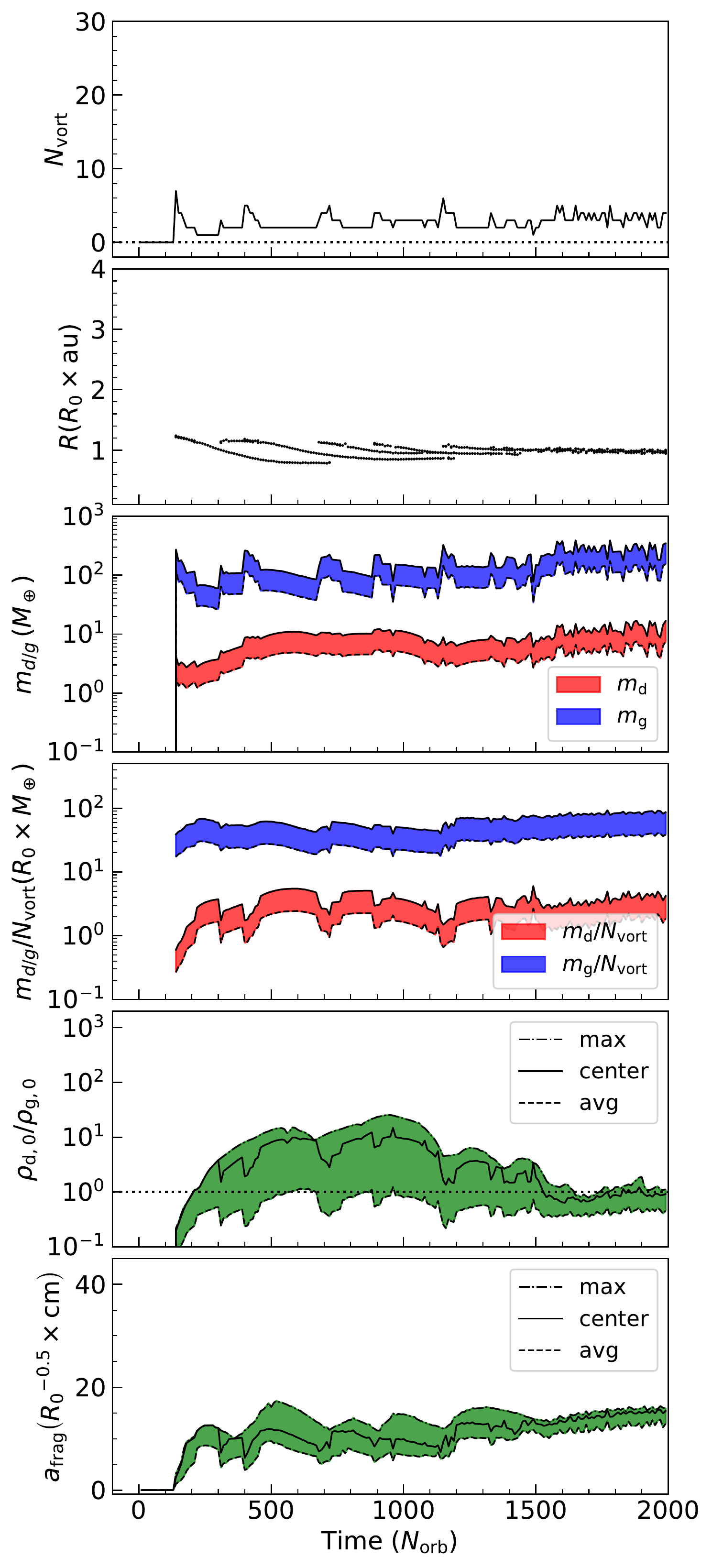} &
  \hspace{-0.5cm}\includegraphics[width=0.25\textwidth]{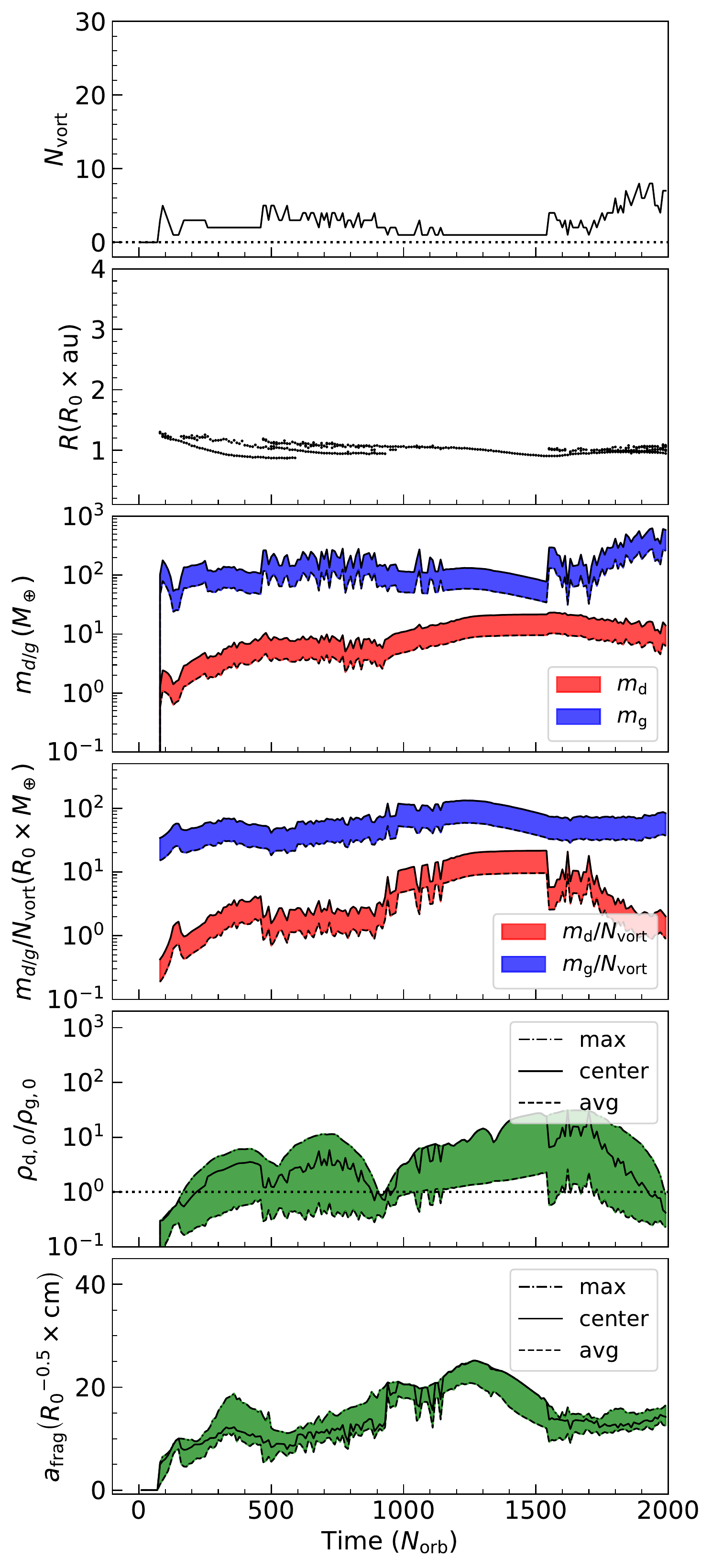} &
  \hspace{-0.5cm}\includegraphics[width=0.25\textwidth]{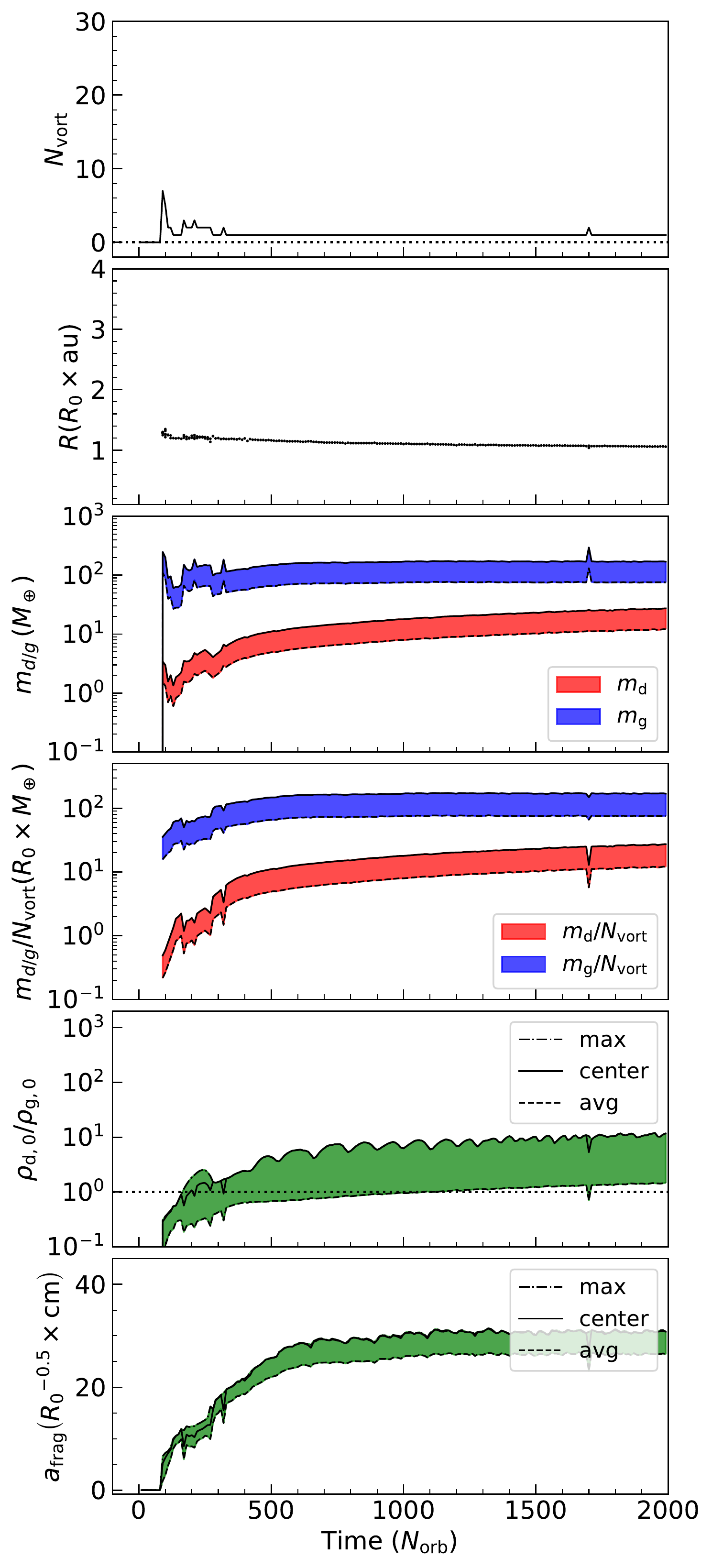}\\
\end{tabular}
\caption{Same as Fig.\,\ref{fig:multi–analysis-2}, but for models B2, C2, D2, and E2, where also multiple vortices develop.}
\label{fig:multi–analysis-2}
\end{figure*}

The simulations conducted in this study correspond to a representative disc, which extends from 0.5~au to 5~au, with the outer boundary of the dead zone set at a radial distance of $R_\mathrm{dze}=R_0=1$~au.
In a typical protoplanetary disc, the dimensions are about an order of magnitude larger, with the outer edge of the dead zone lying at about 15~au.
However, since the self-gravity is not considered in our simulations, the disc can be spatially rescaled. 
In this process, we keep the gas surface density values at 1\,au unchanged, i.e., it corresponds to the aforementioned MMSN models.
With rescaling of the disc size, physical quantities presented in Figs.\,\ref{fig:single–analysis-1}-\ref{fig:multi–analysis-2}  scale with $R_0$ for mass and the fragmentation size scales with $\sim1/\sqrt{R_0}$.
An appropriate scaling for a typical protoplanetary disc is $R_0\simeq10$.
Note that the total mass of the disc will also change with such a rescaling of the disc size.

The strength of a vortex is characterised by  its aspect ratio, $\chi$ (\citealp[see details in e.g.][]{Kida1981,GNG,SurvilleandBarge2015}), such that the vortex reaches its maximum strength at $\chi_\mathrm{min}$ and minimum at $\chi_\mathrm{max}$ value, respectively.
For a large-scale vortex, $\chi$ is obtained by assuming that the density distribution inside the vortex is elliptical and taking the ratio of semi-major to semi-minor axes \citep[][]{Kida1981,Chavanis2000}.
The small-scale vortices are not always elliptical and hence the method of obtaining their aspect ratio is as follows.
The maximum in dust-to-gas ratio is found on the grid, which typically corresponds to the strongest vortex in the disc. 
This is assumed to be the centre of the vortex and centred at this location, contours are drawn at 0.8 of maximum for gas and 0.1 of the maximum for dust.
The ratio of the azimuthal to radial extent of these contours is termed as the aspect ratio.
In Table\,\ref{tbl:RWI}, the minimum value of the aspect ratio is specified, which represents the strongest phase of a vortex.
The last column in this table lists the ratio of total midplane volume density to the Roche density at that radius.
The Roche density is the density required for an incompressible fluid in equilibrium to resist tidal disruption while in a synchronous orbit around a star \citep[e.g.,][]{Chandrasekhar87}. 
This density may be considered as a threshold criterion for gravitational collapse and growth of dust into gravitationally bound clumps.
Note that due to uncertainty in the multiplicative factor, the Roche density is calculated simply as $M_*/R^3$.

\section{Results}
\label{sec:maths} 

\begin{figure*}
    \centering
    \includegraphics[width=1\textwidth]{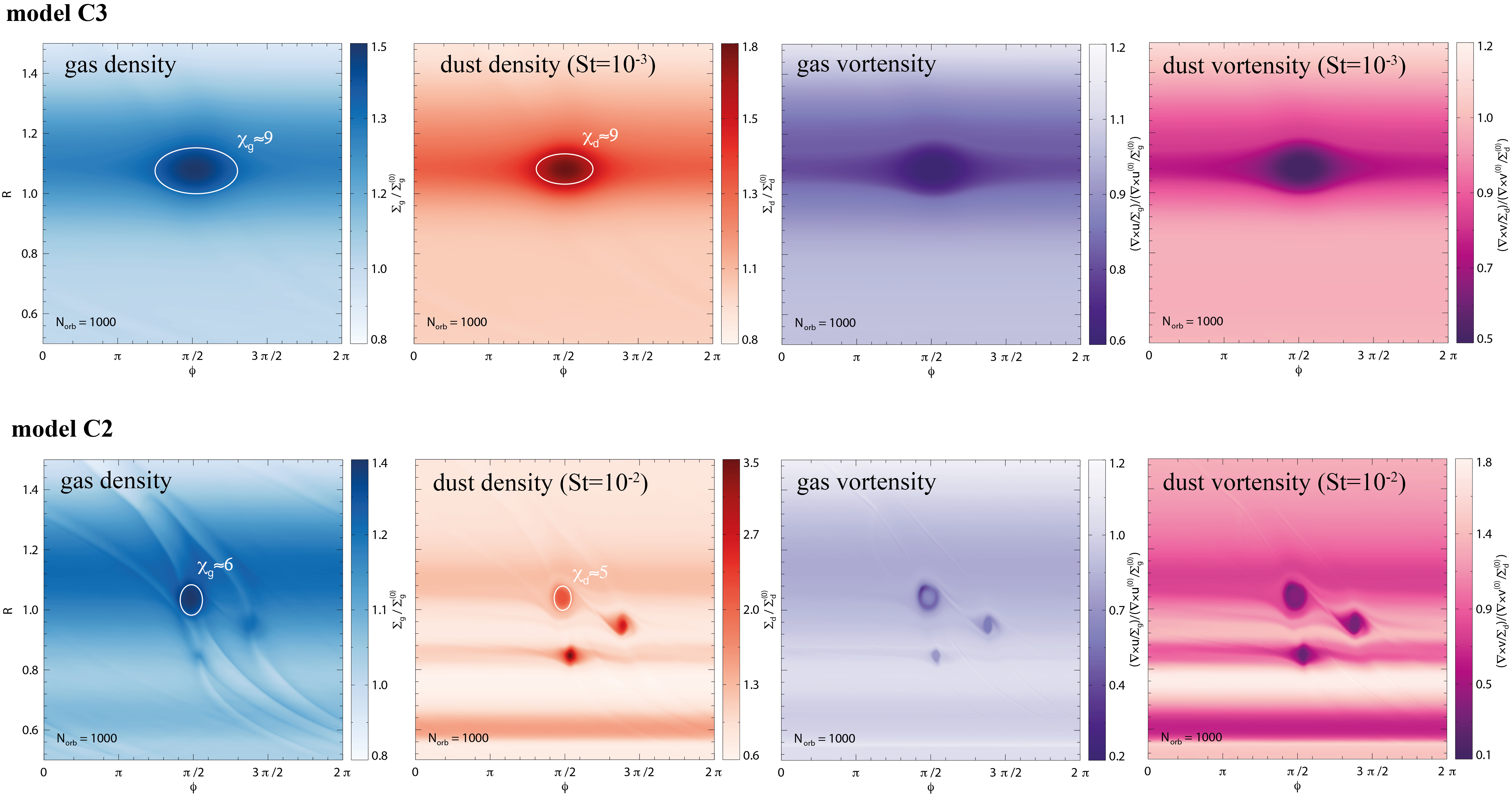}
    \caption{
    An example of the gas and dust surface density distribution, along with respective vortensity fields for the two cases--a single large-scale vortex (model C3 in the upper panels) and multiple small-scale vortices (model C2 in the lower panels).
    The gas and dust components as well as vortensity distributions are normalised with respect to their initial values. 
    The white contours mark the vortex aspect ratio in both gas and dust.}
    \label{fig:comp-C}
\end{figure*}

In summary, although a smooth viscosity transition was assumed at the outer boundary of the dead zone, RWI is excited within 2000 orbits (measured at the outer edge of dead zone) in almost all models, as listed in Table\,\ref{tbl:RWI}.
Exceptions were models A1, C4, and C5.
As inferred earlier, the excitation of RWI can be broadly classified as resulting in one of the two outcomes -- forming either a single large-scale vortex or multiple small-scale vortices.
Fig.~\ref{fig:comp-C} depicts these two outcomes with the help of gas and dust surface density distributions as well as the disc vortensity distributions for models C3 and C2.
The normalised vortensity is calculated as $(\nabla \times \bm{u} / \Sigma_\mathrm{g})/(\nabla \times \bm{u}^{(0)} / \Sigma^{(0)}_\mathrm{g})$ for the gas and $(\nabla \times \bm{v} / \Sigma_\mathrm{d})/(\nabla \times \bm{v}^{(0)} / \Sigma^{(0)}_\mathrm{d})$ for the dust component, respectively.
The vortensity field shows minima associated with the vortices for both the models, which confirms the origin of the vortices in RWI.
Note that all models that exhibit RWI excitation show similar extrema in the vortensity field.
We will come back to this figure after discussing the outcome of the models listed in Table\,\ref{tbl:RWI} in the next two sections.

\subsection{Single large-scale vortex formation}
\label{sec:largescale-vtx}

In the majority of cases, the RWI is excited at an early stage by about 200-500 orbits at $R=1$, i.e., at the location of the outer edge of the dead zone.
The formation of a single large-scale vortex occurs via one of the two evolutionary paths.
In the first case, the ring formed via viscous ring instability fractures into multiple smaller vortices, which promptly merge to form a large-scale vortex.
In the second case, a large-scale vortex slowly appears at $R=1$ at a late stage, after about 1000 orbits.
The excitation of a large-scale vortex can easily be identified by the development of a single maximum in the evolution of azimuthal profiles of gas or dust in Fig.~\,\ref{fig:tromb}. 

Consider the path wherein multiple vortices are formed initially, which rapidly merge into a single large-scale vortex.
The most unstable mode at the excitation of RWI is $m=4$, i.e., four vortices develop initially.
This typically occurs for models with $\mathrm{St}\geq10^{-4}$ (e.g., models A2-A5).
As vortices are formed at slightly different distances and they attract dust deferentially, they approach each other and merge into a single large-scale vortex. A different evolutionary path is taken when the viscosity transition is smoother with $\Delta R_\mathrm{dze}=4H_\mathrm{dze}$ and for a smaller Stokes number of $10^{-5}$ (models A5$^*$, B5$^*$, C3$^*$, and D5$^*$).
This late formation of single vortex can be considered as RWI excitation with an initial mode number $m=1$, which occurs in the dust species that are well-coupled with the gas, in combination with a smoother transition.
Note that model E is an exception, in which RWI excitation occurs via fracture of a VRI ring and at an earlier time at $N_\mathrm{orb}\approx200$.
The larger initial mode of RWI can be observed in Fig.\,\ref{fig:tromb} as a scatter in surface densities, before the merger of individual vortices.

A Rossby vortex is efficient in collecting dust due to the formation of a pressure maximum within its eye.
As the dust is forced to drift towards the local pressure maximum, the azimuthal contrast in the dust is stronger than in the gas.
This is observed in the 2D surface density distribution in Fig.\,\ref{fig:comp-C}, as well as in Fig.\,\ref{fig:tromb}, where the dust patterns are usually more compact as compared to the gas in the evolution profiles.
As seen in Fig.\,\ref{fig:comp-C}, the dust vortensity field also shows greater minima as compared to that for the gas.
This effect weakens if the dust is well-coupled with the gas, resulting in nearly identical dust and gas distributions for $\mathrm{St}\leq10^{-4}$.
After the development of a single large-scale vortex, its azimuthal extension marginally decreases, see, e.g., panels for model A2 in Fig.\,\ref{fig:tromb}.
After reaching a minimum, the vortex starts to grow azimuthally and broaden beyond $\phi>\pi/2$.
The gradual stretching of a vortex occurs due to Keplerian shear, which causes widening of the peak in Fig.~\,\ref{fig:tromb}.
This effect will manifest as an extended horseshoe-like pattern in the evolution of both gas and dust profiles \citep{RegalyVorobyov2017a}.

The gas density distribution of an isolated vortex has been shown to be elliptical in shape \citep[][]{Kida1981,Chavanis2000}. 
The aspect ratio is a measure of its strength such that a lower value indicates higher strength in terms of vortensity (\citealp[][]{Kida1981,GNG,SurvilleandBarge2015}). 
A lower value of $\chi$ represents the strongest phase of the vortex, which is listed for both dust and gas for each simulation in Table \ref{tbl:RWI}.
In general, $\chi_\mathrm{min}$ in gas and dust are the same for $\mathrm{St}\leq10^{-4}$. However, the vortex becomes stronger, i.e., $\chi_\mathrm{min}$ is smaller in dust for $\mathrm{St}\geq10^{-3}$ than in gas.
Except for model set A, wherein the gaseous component shows small number of $\chi_{\rm min}$.
Note that for such a low aspect ratio,  vortices might be strongly unstable in 3D, see, e.g., \cite{LesurPapaloizou2009}.
Note that in models A5*, B5*, C3* (where $\Delta R_\mathrm{dze}=4H_\mathrm{dze}$ is assumed), the vortex can not reach its strongest phase by the end of simulation because of the late excitation of RWI.
Considering the last column in Table \ref{tbl:RWI}, the central density of a large-scale vortex is much less than Roche density. This suggests that these vortices can not exhibit gravitational collapse, as the self-gravity is insufficient to overcome the tidal disruption by the star.

We now discuss the evolution of large-scale vortices in terms of their temporal evolution as shown in Figs.~\ref{fig:single–analysis-1} and \ref{fig:single–analysis-2}. 
For each subfigure, the uppermost panel shows the distance of vortex centre from the central star and the second panel shows the mass of dust and gas accumulated in the vortex. 
Considering the first panel for all models, a vortex drifts only marginally towards the star and stays close to the location of the viscosity transition near 1$R_0$.
Note, however, that the long-term evolution of viscosity transition itself was not modelled here.
Is seen that vortices only slightly drift toward the star and by reaching their maximum strength the direction of the drift changes due to vortex dynamics and erosion.
Analysis of the evolution of the mass collected by the vortices shows that a large-scale vortex can accumulate several hundred $R_0\times M_\oplus$ gas, irrespective of the Stokes number. 
The gas mass accumulated in a vortex is similar for all models.
This value corresponds to about twice the mass of Jupiter in gas, assuming $R_0\simeq10$ scaling. 
The dust mass accumulated inside the vortex is inversely proportional to the Stokes number, e.g., as observed in models A2, A3, and A4.  
The amount of dust collected is between 1 and $10\, R_0\times M_\odot$.
With the $R_0\simeq10$ scaling, these values correspond to 10 to 100 Earth masses.
For model A2, with $\mathrm{St}=10^{-2}$ the collected dust mass can reach $10\,R_0\times M_\odot$, while for other models less dust is accumulated. By applying $R_0\simeq10$ scaling, model A2 accumulates about a hundred or ten Earth masses.

The last two panels in Figs.~\ref{fig:single–analysis-1} and \ref{fig:single–analysis-2} show the ratio of central dust-to-gas volume density and the fragmentation size of the dust particles, respectively.
We found that the larger the Stokes number, the larger the dust-to-gas density ratio (e.g., models A2, A3, and A4).
In general, the dust-to-gas density ratio remains low between  $10^{-2}$ and $10^{-1}$.
A small spread between the maximum and minimum bounds of this ratio indicates a weak vortex, which is unable to concentrate the dust efficiently or a strong coupling between the dust and gas components.
However, in models A1 and A2 the dust-to-gas density ratio can exceed unity and thus, excitation of the streaming instability is possible.
Note that the highest dust density is almost (except in model A1) always at the vortex centre as the max and centre values overlap.

With regards to the growth of solid species, we found that $a_\mathrm{frag}$ lies in the range of $20-40 \times (1/\sqrt{R_0})$\,cm by about 1000 orbits, independent of the model details. With the $R_0\simeq10$ scaling of a disc, it corresponds to $6-12$\,cm. The maximum value of $a_\mathrm{frag}$ is measured at the vortex centre, which implies that the dust growth is most efficient at this location.
A notable observation for the case E of model parameters (models E3-E5) is that $a_\mathrm{frag}$ initially increases and then gradually decreases as the vortices evolve.  
The gradual decrease in $a_\mathrm{frag}$, in general, is caused by a marginal decrease in gas density as the simulations proceed.
This effect is magnified for the model set E due to its strong dependence on the gas surface density. 
Finally, we emphasise that all other physical quantities presented in Figs.~\ref{fig:single–analysis-1} and \ref{fig:single–analysis-2}, do not change significantly beyond 1000 orbits. 
There are three exceptions--models B5*, C3*, D5*--wherein the vortex is not fully-fledged by the end of simulation due to the late excitation of RWI.

At this point we mention the anomalous behaviour of model A1. 
As this simulation progresses, the dust distribution near the dead zone edge resembles azimuthally elongated streaks and not well-formed vortices.
In Fig.~\ref{fig:single–analysis-1}, the quantities are plotted at the location of the highest dust concentration. 
The rapid fluctuation of this location as well as properties associated with it reflects the dynamic nature of the dust-gas system.
We hypothesise that such evolution is caused by strongly concentrated dust. Since we do not model formation of planetesimals, the concentrated dust remains in the disc and manifests as elongated streaks. 
Such streaks are transiently observed in other models with a large Stokes number, although to a much lesser extent.  

\subsection{Multiple vortex formation}
\label{sec:multi-vtx}

The nature and evolution of the small-scale vortices are  remarkably different and much more complex as compared to the large-scale vortices.
Small-scale vortices are formed when the dusty rings developed via the viscous ring instability become Rossby unstable and typically breaks up with a large azimuthal mode number ($m \simeq 10-20$). 
Although the resulting vortices show some merger, prompt formation of a single large-scale vortex does not occur.
Long-term survival of multiple vortices in the disc opens up the possibility of complex vortex-vortex interactions.
An individual vortex typically gathers dust, produces large-scale spiral waves in the dust and migrates inward.
The constructive interference between such spiral waves in certain cases may give rise to a cascading effect such that a new generation of vortices is induced at a different radius in the disc.
It is also possible that multiple rings first form via viscous ring instability, which in turn become Rossby unstable.
Thus, depending on the model parameters, the process of self-sustaining vortices can repeat multiple times and cascade throughout the disc, see more details of these phenomena in \cite{Regalyetal2021}.
The $\delta\Sigma$ profiles show non-uniformity and scattered features in Fig.\,\ref{fig:tromb}, e.g., in models A1, A2, B1, B2, C1, C2, D1, D2, E1, and E2. 
These features reflect the presence of multiple vortices in the disc as well as possible shift of the maximum density on the grid from one vortex to another.
Thus, we need additional methods of analysis when the disc shows formation of multiple vortices.

Figs. \ref{fig:multi–analysis-1} and \ref{fig:multi–analysis-2} show the evolution of several quantities for the models that develop multiple vortices.
These indicators are useful for gaining insight into the bulk behaviour of the vortices formed within the disc and are obtained using methods described in Section \ref{subsec:moa}.
The first panel of each subfigure in Figs.~\ref{fig:multi–analysis-1} and \ref{fig:multi–analysis-2} shows the number of vortices identified in the disc ($N_\mathrm{vort}$), while the second panel shows the radial position of each vortex at a given time.
Vortex-vortex mergers result in both the initial steep decline in the number of vortices and  eventual gradual decline.
Due to a limited vortex merging, $N_\mathrm{vort}$ stays between 10 and  1 by the end of simulations.
With regards to the radial position of small-scale vortices, a tendency to drift inwards is observed.
However, as the vortices reach the global pressure maximum at the viscosity transition near $1 {\rm R_0}$ and the inward drift stops almost completely. 
Thus, the small-scale vortices tend to get trapped at the viscosity transition at the outer boundary of the dead zone.
For two of the models, C1 and D1 (Fig.\,\ref{fig:multi–analysis-1}) vortices can form well outside the dead zone.
This phenomenon is similar to vortex cascade described in \citet{Regalyetal2021}, wherein formation of multiple generations of vortices occurs due to constructive interference of spiral waves created by vortices that are present at a given radius.

Consider the next two panels of Figs.~\ref{fig:multi–analysis-1} and \ref{fig:multi–analysis-2}. The third panel shows the cumulative mass of both dust and gas accumulated in all the vortices present in the disc at a given time ($m_{\rm d}$ and $m_{\rm g}$), while the fourth panel shows the average mass accumulated by an individual vortex.
Note that the upper and lower bounds are obtained from the upper and lower estimates for an MMSN disc (see Section \ref{subsec:moa}).
For the models that show formation of multiple vortices, the upper bound of material collected by an individual vortex can exceed $100\times R_0\,M_\oplus$ for the gas very quickly, which corresponds to about twice the mass of Jupiter for $R_0\simeq10$ scaling.
In all multiple vortex models, the amount of material collected by the largest (strongest) vortex can reach $100\times R_0\,M_\oplus$ for the gas very quickly, which corresponds to a couple of Jupiter mass for $R_0\simeq10$ scaling.
The dust content of a vortex typically increases rapidly in the beginning and flattens out with time. 
In models B1, C1, D1, and E1, a vortex can collect over $10\times R_0\,M_\oplus$ of dust, corresponding to a value of $100\,M_\oplus$ for $R_0\simeq10$ scaling by the end of the simulation. 
This means that the dust-to-gas mass ratio inside an individual vortex is enhanced by over ten times in general compared to the initial value in the disc.
For models B2, C2, D2, and E2, the dust collected by a vortex is below $10\times R_0\,M_\oplus$, implying a marginal increase over the initial dust-to-gas mass ratio.

With regards the second to last panel in these figures showing the dust-to-gas density ratio, the models can be grouped in the same two classes.
In models shown in Fig.\,\ref{fig:multi–analysis-1}), the dust-to-gas density ratio increases continuously and can stay well above unity, while the maximum can approach 1000.
In models B2, C2, and D2 (shown in Fig.\,\ref{fig:multi–analysis-2}) the oscillation of dust-to-gas density reflect destruction of vortices and collection of dust the new ones.
In models B2, C2, and D2 (shown in Fig.\,\ref{fig:multi–analysis-2}), the dust-to-gas density ratio shows cyclic oscillations due to the occurrence of vortex cascade. Within these oscillations, the minima reflect destruction of vortices as they evolve and the maxima correspond to formation of the next generation of vortices.   
Another observation of interest is that the maximum dust-to-gas density ratio is not coincident with the average value at the vortex centres.
This reflects a differential gathering of dust, i.e., one vortex gathering more dust as compared to the rest.
A secondary cause of this difference is the local vortex dynamics, which result in large, asymmetric vortices which show off-centre accumulation of the dust.  

Similar to the dust-to-gas density ratio, the maximum size that the solid can reach increases monotonically with time. 
Similar to the behaviour of the dust-to-gas ratio, models B2, C2, and D2, show a trend of varying $a_\mathrm{frag}$ because of vortex cascade.
The dust can typically grow up to $a_\mathrm{frag}\approx20\times(1/\sqrt{R_0})$\,cm, which is about half the size that is achieved in the case of large-scale vortices.
For $R_0\simeq10$ scaling this corresponds to about 6~cm.
As seen in Table \ref{tbl:RWI}, the central density in small-scale vortices can exceed Roche density, but only for the case of large Stokes number.  
Density exceeding Roche value implies that the collected dust may undergo gravitational collapse, forming more massive, gravitationally bound structures.

\section{Discussion}

\begin{figure*}
    \includegraphics[width=\textwidth]{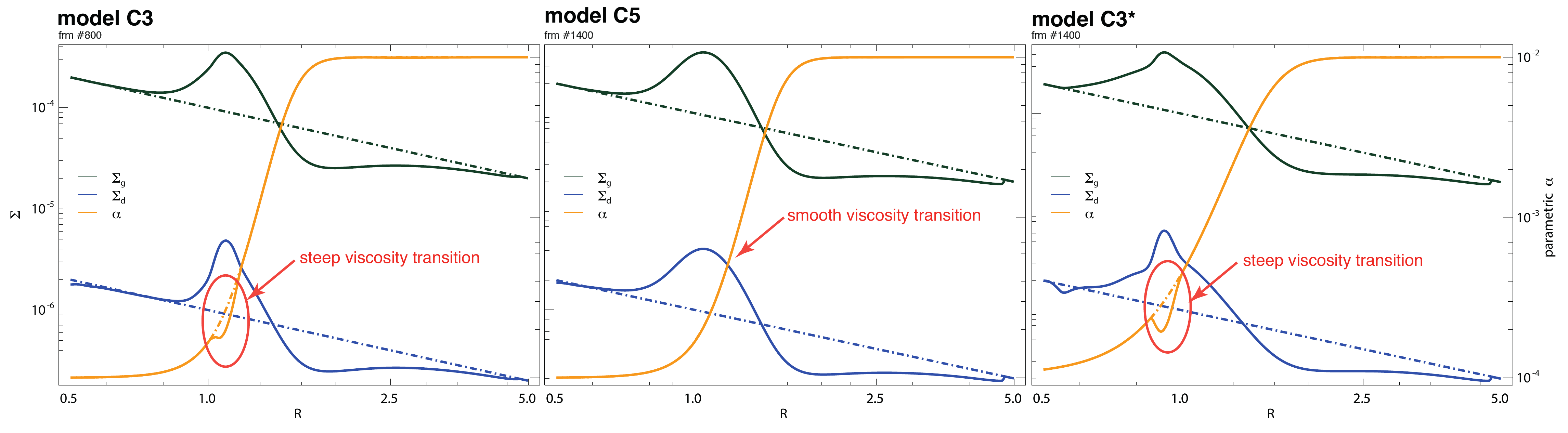}
    \caption{Radial profiles of the azimuthally averaged surface mass density of the gas ($\Sigma_\mathrm{g}$) and dust ($\Sigma_\mathrm{d}$) and the value of $\alpha$ parameter in three models C3, C5, and C3*. The profiles are calculated from snapshots taken by 800, 1400, and 1400 orbits at $R=1$, respectively. Dot-dashed curves show the initial profiles. $\alpha$ profile has a steep region in models where the local dust-to-gas density ratio grows beyond the initial value of 0.01. RWI is eventually  RWI excited in models C3 and C3*.}
    \label{fig:profile1}
\end{figure*}

It is well-established that RWI excitation is inhibited if the half-width of the viscosity transition exceeds the local pressure scale height by twofold \citep{Lyraetal2009,Regalyetal2012}. 
This is because of the mismatch in mass transfer rate due to varying viscosity is insufficient to sustain a strong  pressure gradient.
However, as we showed in the previous sections, with a dust-dependent prescription for viscosity, RWI excitation occurs even when a smooth transition is assumed. 
We explain the formation of vortices in our simulation with the help of Fig.\,\ref{fig:profile1}.
The figure shows azimuthally averaged profiles of gas and dust surface densities as well as the effective $\alpha$-parameter in models C3, C5, and C3*. 
An enhancement in gas surface density forms in all models due to the mismatch of the accretion rate near the viscosity transition. 
Since such a profile creates a maximum in gas pressure, it collects dust and the local dust-to-gas mass ratio increases. 
According to the parametric-$\alpha$ prescription, see Equation~\ref{eq:alpha}, this increasing concentration of dust results in a decrease in the viscosity.
The decreased viscosity leads to further enhancement of the gas surface density and a positive feedback loop ensues.
As a result, RWI is soon excited despite the initially smooth viscosity transition, with $\Delta R_\mathrm{dze}=(2+2/3)H_\mathrm{dze}$ (model C3) or even $R_\mathrm{dze}=4H_\mathrm{dze}$ (model C3*).

The above-described mechanism works effectively in almost all cases for $\mathrm{St}<10^{-2}$ models. However, in C4 and C5 models, the local enhancement in the dust density is weak, namely the density peak in dust is too smooth to effectively sharpen the viscosity transition, as seen in the middle panel of Fig.~\ref{fig:profile1}. This is because of that the dust is well coupled to the gas as $\mathrm{St}=10^{-4}$ and $10^{-5}$, and in model C, the parametric-$\alpha$ prescription assumes $\Phi_\mathrm{g}=1$, for which case the density enhancement in gas is against the local viscosity depression. Note that in these models RWI is not excited. In case of $\mathrm{St}\geq10^{-2}$ RWI excitation and subsequent vortex cascade occur in the same fashion as was identified in \citet{Regalyetal2012}.

Although the positive feedback due to dust-dependent $\alpha$-parameter described above acts in all models, the exact qualitative outcome, e.g., if small-scale or large-scale vortices form,  depends on the model parameters. 
These parameters are the exponents $\phi_{\rm g}$ and $\phi_{\rm d}$, which determine the attenuation of the disc viscosity as well as the Stokes number, which specifies the coupling properties of the dust species.
The process of dust diffusion tends to smear out sharp gradients in dust surface density and weaken a vortex. 
However, the process of dust drift causes its migration towards a local maximum in gas pressure and enhances its concentration.
Ultimately the formation of vortices in a particular model is determined by the balance of these two processes, which work in the opposite direction.
The dust particles with a relatively large Stokes number, e.g., models with $\mathrm{St}>10^{-2}$, move rapidly towards the local pressure maxima, while they also suffer less diffusion (see Eq. \ref{eq:diffusion}).
Since sustenance of small-scale vortices will require preservation of a strong gradients of dust surface density, these conditions are satisfied only at a relatively large Stokes number.
On the other hand, small-sized dust is well coupled to the gas motion and drifts slowly within the disc. 
Such dust particles cannot maintain gradients required to sustain small-scale vortices and thus, only a single large-scale vortex may form. 
This trend for Stokes number is indeed observed in our simulations in Table \ref{tbl:RWI}. 
Thus, only a bimodal outcome is possible for the fate of evolving vortices, wherein either a cascade of multiple small-scale vortices occurs or a single large-scale vortex is formed.
The phenomenon of differential drift also results in increased Roche densities for large Stokes number as listed in Table \ref{tbl:RWI}.
If diffusion process dominates the dust dynamics, e.g., model C5 in Fig.~\ref{fig:comp-C}, the dust is unable to concentrate efficiently at the pressure maximum and formation of a vortex is suppressed.
If the dust drift is particularly strong, this may result in evolution similar to model A1.
In such a case, a relatively large extent of the disc is prone to streaming instability. Since we do not resolve streaming instability or model planetesimal formation, the dust remains in the disc and this manifests as elongated streaks.
Note that diffusion coefficient also depends on the local viscosity, which in turn depends on the exponents $\phi_{\rm g}$ and $\phi_{\rm d}$.
The exact outcome of a particular combination can only be determined via conducting self-consistent simulations.

Both small-scale and large-scale vortices typically originate near the smooth dead zone edge and although some inward migration is observed, they do not migrate into the dead zone. 
The vortex migration is most notable in models C1 and D1.
Considering these two cases with $R_0=10$ scaling, the radial migration speed of a vortex when it is farther away beyond 20 au is approximately 3.3-4.2 ${\rm m\,s^{-1}}$. Closer to the dead zone edge at 10 au, the speed is slower at about 1.4-2.1 ${\rm m\,s^{-1}}$. 
For comparison, the drift velocity of dust with ${\rm St}=0.1$ is $3.3 {\rm m\,s^{-1}}$ at 20 au and 4.7 ${\rm m\,s^{-1}}$ at 10 au for an unperturbed $\alpha$-disc. 
Thus, we can conclude that away from the dead one edge, the vortex drift is almost entirely due to the dust drift. 
The simulations show that in the vicinity of the outer edge of the dead zone, the pressure bump formed is sufficient to slow down and eventually completely halt the inward migration of vortices in all cases. 
This loitering of the vortices at the dead zone edge potentially has significant consequences for planet formation. 
Note that since the simulations do not consider disc self-gravity, the role of vortex-disc interactions is minimal and the migration is unlike the type I migration of planets.

Another observation of interest is that the dust collected in vortices is typically off-centre and this is most notable in the case of small-scale vortices.
Similar phenomenon of off-centre dust concentration has been reported by \cite{Hammer2019}. In their case, presumably by repeated perturbations from the spiral arms of a planet that is present in the disc.
For the small-scale vortices in our simulations, the perturbations from other vortices may cause similar asymmetrical dust concentrations.
Although detailed behaviour of both the small and large scale vortices in our simulations warrants further exploration, which will be carried out in the future.

Merger of vortices is typically observed for the small-scale vortices shortly after their formation, although in most cases, not all vortices in the disc merge to form a single vortex.
Once a vortex is formed, it quickly starts attracting dust from its surrounding.
A small difference in accumulated dust as compared to its peers may nudge a vortex inwards in the disc.
This differential dust gathering, in addition to the fact that small-scale vortices have strong local dust concentrations as well as a small size, may explain how several vortices often co-evolve in the disc. 
A large-scale vortex may form in the disc in two ways. 
Several smaller vortices typically with $m=4$ are formed, which shortly merge into a single large-scale vortex (e.g., D3, D4, D5) or a large-scale vortex gradually emerges at the pressure maximum (e.g., B5*, D5*).
We hypothesise that the exact outcome depends on the difference between the relative strength of the dust diffusion as compared to its drift, with the former favouring gradual formation of a large-scale vortex.

Figs.~\ref{fig:single–analysis-1}-\ref{fig:multi–analysis-1}, some general trends can be observed. 
The small-scale vortices are always much stronger than the large-scale vortices with respect to the midplane ratio of volume density, $\rho_d/\rho_g$. The value of $\rho_d/\rho_g$ remains below unity for all large-scale vortices, except A2, where it remains constant at unity. 
On the other hand, for small-scale vortices, this ratio exceeds unity, and in some cases the maximum values approach 1000.
The origin of this disparity can be traced back to the balance between the dust drift and its diffusion. 
Small-scale vortices can be sustained only in the cases where the dust drift dominates, and this results in strong concentrations of dust and an enhanced dust-to-gas mass ratio.

An opposite trend in the dust fragmentation size is observed, where $a_\mathrm{frag}$ is approximately twice as large inside large-scale vortices as compared to the small-scale vortices.
The fragmentation size scales as $a_\mathrm{frag}\sim \Sigma_\mathrm{g}/c_\mathrm{s}^2$ according to Eq.~(\ref{eq:afrag}).
We found no significant difference in the temperature inside the two types of vortices, which implies that the local sound speed is not responsible for observed disparity.
However, the gas density in the large-scale vortices is typically about twice as large as compared to the small-scale vortices, which explains the discrepancy in $a_\mathrm{frag}$ for the two scenarios.

It is known that accretion discs might be a subject of axially symmetric pulsational instability similar to stellar oscillations \citep{Kato1978}. Since the thermal energy is supplied in part by the viscous dissipation of shear motion and the viscosity depends on the local gas compression via Equation\,(\ref{eq:alpha}), any oscillations in the disc can get amplified.
This phenomenon is known as viscous overstability and \citet{LatterOgilvie2006} give an analytical criterion for its occurrence in accretion discs. 
Since the disc viscosity in our models is confined between $10^{-4}\leq\alpha\leq10^{-2}$ and the disc vertical thickness is around $H/r\simeq0.05$, viscous overstability does not occur on the long-wavelength limit. 
However, for short wave-length limit we have to analyse Equation\,(14) of \citet{LatterOgilvie2006} with the assumption of
\begin{equation}
\frac{\mathrm{d}(\nu\Sigma_\mathrm{g})}{\mathrm{d}\Sigma_\mathrm{g}}=\nu(1+\phi_\mathrm{g})\left(\frac{\Sigma_\mathrm{g}}{\Sigma_\mathrm{g}^{(0)}}\right)^{\phi_\mathrm{g}}.
\end{equation}
In the limit of zero bulk viscosity for the gas\footnote{Being negligible small the gas bulk viscosity, we neglect it. Moreover, at the applied grid resolution, the magnitude of the numerical viscosity of the FARGO algorithm is also negligible compared to the assumed $\nu=\alpha c_\mathrm{s}^2/\Omega$ effective viscosity.}, the viscous overstability occurs on the wavelength 
\begin{equation}
\frac{1}{k}\gtrsim\frac{c_\mathrm{s}}{\Omega}\left(\frac{4}{9(1+\phi_\mathrm{g})(\Sigma_\mathrm{g}/\Sigma_\mathrm{g}^{(0)})^{\phi_\mathrm{g}}-7}\right)^{1/2},
\end{equation}
where we assume that $\nu\ll c_\mathrm{s}^2/\Omega$. 
For an unperturbed disc ($\Sigma_\mathrm{g}/\Sigma_\mathrm{g}^{(0)}=1$), the above criterion gives  $\sqrt{2}H$ for $\phi_\mathrm{g}=0$, i.e., viscous overstability would be active over the disc vertical scale height. For $\phi_\mathrm{g}=1$, we get $\sqrt{4/11}H$, i.e., our results may be affected by the viscous overstability. For the case of $\phi_\mathrm{g}=1$, there is no valid solution for $1/k$, meaning no viscous overstability occurs.

Finally, we mention some constraints on the detectability of vortices formed in our dust-dependent $\alpha$ model. 
Large-scale vortices in protoplanetary discs have been already detected in the millimetre wave-length observation by SMA or ALMA radio interferometers \citep[e.g.,][and references therein]{Regalyetal2012,Regalyetal2017}. 
On the other hand, a small-scale vortex in our model typically extends about 3 au assuming $R_0=10$\,au scaling. At 100\,pc, this subtends 0.03 arcseconds. Since ALMA in C–10 configuration can resolve up to 0.018–0.012 arcseconds\footnote{See Cycle 9 Proposer’s Guide at https://almascience.nrao.edu for details.} in band 6 and 7, respectively, these vortices are theoretically detectable. However, if such vortices indeed form in an early protoplanetary disc, the dust within them may rapidly grow into planetesimals/protoplanets on a very short timescale, making direct observations difficult.

\section{Conclusions}
  
In this paper, we present the results of numerical experiments that investigate the excitation of Rossby vortices in protoplanetary discs at the outer edge of the dead zone. 
Canonically, an abrupt transition in viscosity is required for the excitation of Rossby wave instability, with a half-width no more than twice the local pressure scale height (i.e., $\Delta R_\mathrm{dze}< 2 H_\mathrm{dze}$).
However, the viscosity transition at the outer edge of the dead zone has been shown to be smooth and gradually varying.
We conducted hydrodynamic simulations using a parametric–$\alpha$ model, wherein the MRI efficiency depends on the local concentration of dust as well as gas due to adsorption of charged particles on the grain surface.
With such a dust-dependent $\alpha$ formulation, with the viscosity being a function of the local concentration of dust as well as gas, RWI can be excited in most cases despite a smooth viscosity transition.

The dust-gas coupled simulations were conducted in the thin-disc limit with adiabatic disc thermodynamics.
Four cases of dust-dependent parametric-$\alpha$ models were investigated (see Table\,\ref{tbl:cases} for details).
The dust component was assumed to have a fixed Stokes numbers in the range of $10^{-5}\leq \mathrm{St}\leq10^{-1}$.
The considered viscosity transitions were smooth, specifically, with a width of $\Delta R_\mathrm{dze}=(2+2/3)H_\mathrm{dze}$ and $\Delta R_\mathrm{dze}=4H_\mathrm{dze}$.
We found that RWI is excited in almost all models, resulting in formation of anticyclonic vortices.
The excitation of RWI, despite smooth viscosity transition, can be explained by the local steepening of $\alpha$ due to dust enhancement and a positive feedback cycle between dust accumulation and a reduction in $\alpha$-parameter.
The non-excitation of RWI in two of the cases can be explained by a weak dust enhancement due to strong coupling of the dust species ($\mathrm{St}\leq 10^{-4}$) and a positive dependence of $\alpha$ on the gas density ($\phi_\mathrm{g}=1$). 

Two distinct outcomes of RWI excitation were identified -- formation of a single large-scale vortex or generation of multiple small-scale vortices.
Table\,\ref{tbl:RWI} lists the excitation outcome for the considered combinations of parametric-$\alpha$ model and Stokes number.
In summary, the exact outcome of a simulation is determined by the balance between dust diffusion and its drift towards the local gas pressure maximum.
A large-scale vortex develops when the dust is well coupled to the gas ($\mathrm{St}\leq10^{-3}$) or when $\alpha$ is independent of dust concentration (model set A). 
In such a case, the dust particles cannot maintain gradients required to sustain small-scale vortices.
The small-scale multiple vortex scenario occurs only for less coupled dust species, with $\mathrm{St}\geq10^{-2}$.
During their evolution, small-scale vortices may merge, however, formation of a single large-scale vortex is avoided.

In a canonical MMSN disc, the gas mass accumulated in a typical large-scale vortex is approximately twice the mass of Jupiter.
For $\mathrm{St}=10^{-1}$ and $\mathrm{St}=10^{-2}$ a large-scale vortex can collect about 100 and 10 Earth mass of solid material, respectively.
The dust-to-gas density ratio generally remains below unity and typically increases with an increases with the Stokes number. 
For model A2 with $\mathrm{St}=10^{-2}$, the dust-to-gas density ratio exceeded unity, in which case the streaming instability can be excited.
The total midplane density exceeds Roche density only for the case of largest Stokes number, ${\rm St}=10^{-1}$, indicating that a direct gravitational collapse is also feasible.
With typical assumptions of dust properties, the dust can grow within a large-scale vortex to a fragmentation size of about 6-12~cm.

In the case of formation of multiple small-scale vortices in the disc, an individual 
vortex can collect about 1 to 10 Earth masses of solid material.
The dust-to-gas density ratio grows well above unity in general, and can even reach a hundred in certain cases. 
As a result, the criterion for streaming instability is always satisfied in small-scale vortices.
As compared to a large-scale vortex, the gas density is not enhanced by a large margin by a small-scale vortex.
As a result, the maximum size that the dust can reach is about 6~cm, half of that found in large-scale vortices.

It is shown in \citet{Regalyetal2021} that small-scale vortices developed with a dust-dependent parametric-$\alpha$ model are subject to relatively fast inward migration.
However, in this study we found that the migration of small-scale as well as large-scale vortices halts at the dead zone edge.
The trapping of small-scale vortices at the outer dead zone edge is particularly noticeable in models C1 and D1 (see Fig.\,\ref{fig:multi–analysis-1}).

The formation of vortices via Rossby instability at the smooth outer dead zone edge in protoplanetary discs occurs for a wide range of parameters in the dust-dependent $\alpha$ models as well as Stokes numbers.  
The resulting large- as well as small-scale vortices get trapped at the dead zone edge and remain stable over hundreds or thousands of orbits.
Thus, the meter size barrier can be overcome within the vortices to form planetesimals and planetary embryos.
The tens of Earth masses in cumulative solid material collected by the vortices is sufficient to form planetary systems similar to our own solar system \citep{Weidenschilling77}.  
Thus, we conclude that such vortices formed at the outer dead zone edge of protoplanetary discs act as planetary nurseries, providing ideal environment for dust growth into planetesimals and beyond.

\section*{Acknowledgements}

We thank the anonymous referee of improving the quality of the manuscript. The project was supported by the Hungarian OTKA Grant No. 119993. R.Zs acknowledges helpful discussions with V. Fröhlich. . K.K. acknowledges support from NSERC of Canada. D. T.-N. acknowledges the financial support of the Lend\"ulet Program  of the Hungarian Academy of Sciences, and project No.  LP2018-7/2021 and the KKP-137523 'SeismoLab' \'Elvonal grant of the Hungarian Research, Development and Innovation Office (NKFIH).

\section*{Data availability}
The data underlying this article obtained with GFARGO2 code can be shared at a reasonable request to the corresponding author.




\bibliographystyle{mnras}
\bibliography{references} 








\bsp	
\label{lastpage}
\end{document}